\documentclass{optica-article}

\journal{opticajournal} % for journals or Optica Open

\articletype{Research Article}

\usepackage{lineno}
\usepackage{siunitx}
\usepackage[normalem]{ulem}

\usepackage[utf8]{inputenc}
\usepackage[T1]{fontenc}
\usepackage{graphicx}
\usepackage{float}
\usepackage{siunitx}
\usepackage{epstopdf}
\usepackage{caption}
\usepackage{geometry}
\usepackage{mathtools}
\usepackage{hyperref}
\hypersetup{colorlinks=true,       % false: boxed links; true: colored links
    linkcolor=red,          % color of internal links (change box color with linkbordercolor)
    citecolor=blue,        % color of links to bibliography
    filecolor=magenta,      % color of file links
    urlcolor=cyan           % color of external links
    }
\usepackage{subcaption}
\usepackage[english]{babel}
\usepackage{soulutf8}

\DeclarePairedDelimiterX{\norm}[1]{\lVert}{\rVert}{#1}

\renewcommand{\eqref}[1]{\text{Eq.\ (}\ref{#1}\text{)}}

\begin{document}

\title{Evaluating probabilistic and data-driven inference models for fiber-coupled NV-diamond temperature sensors}
%\title{Evaluating the impact of probabilistic and data-driven inference models on uncertainties of fiber-coupled NV-diamond temperature sensors}
\author{Shraddha Rajpal,\authormark{1,*} Zeeshan Ahmed,\authormark{2} and Tyrus Berry \authormark{3}}

\address{\authormark{1}School of Mathematical and Statistical Sciences, Clemson University, Clemson, SC 29634, USA\\
\authormark{2}Sensor Science Division, Physical Measurement Laboratory, National Institute of Standards and Technology, Gaithersburg, MD 2010 , USA\\
\authormark{2}Department of Mathematical Sciences, George Mason University, Fairfax, VA 22032,USA}

\email{\authormark{*}zeeshan.ahmed@nist.gov}

% use {asbstract*} to suppress the copyright line. Copyright information will be added in production

\begin{abstract}
We evaluate the impact of inference model on uncertainties when using continuous wave Optically Detected Magnetic Resonance (ODMR) measurements to infer temperature. Our approach leverages a probabilistic feedforward inference model designed to maximize the likelihood of observed ODMR spectra through automatic differentiation. This model effectively utilizes the temperature dependence of spin Hamiltonian parameters to infer temperature from spectral features in the ODMR data. We achieve prediction uncertainty of $\pm$ 1  K across a temperature range of 243 K to 323 K.  To benchmark our probabilistic model, we compare it with a non-parametric peak-finding technique and data-driven methodologies such as Principal Component Regression (PCR) and a 1D Convolutional Neural Network (CNN).  We find that when validated against out-of-sample dataset that encompasses the same temperature range as the training dataset, data driven methods can show uncertainties that  are as much as 0.67 K lower without incorporating expert-level understanding of the spectroscopic-temperature relationship.  However, our results show that the probabilistic model outperforms both PCR and CNN  when tasked with extrapolating beyond the temperature range used in training set, indicating robustness and generalizability. In contrast, data-driven methods like PCR and CNN demonstrate up to ten times worse uncertainties when tasked with extrapolating outside their training data range.
\end{abstract}

\section{Introduction}
Temperature is a critical parameter in various aspects of modern life, including manufacturing processes \cite{woo2009adaptive}, medical procedures, and environmental control of residential and commercial spaces. To meet such disparate measurement needs, a variety of temperature sensors have been developed. Although these devices vary greatly in their cost, size, weight and complexity, they almost all rely on well-established measurements of transport properties to infer temperature. Legacy technologies like platinum resistance thermometers and negative temperature coefficient thermistors have been relied upon for over a century to provide accurate and reproducible measurements over a broad range of temperature\cite{price1959platinum,strouse2008standard, review2022}. However, these sensors are prone to drift and require frequent re-calibrations to ensure high accuracy in critical use-cases resulting in increase cost of sensor ownership.

In recent years, there has been a growing interest in developing alternative sensor technologies that can overcome the limitations of traditional technologies. The past decade has seen a burst of activity in nanophotonics\cite{xu2014ultra}, quantum optomechanics\cite{purdy} and noise thermometry\cite{jnt}. These technologies leverage telecom industry's vast economies of scale along with  precision measurement expertise developed for frequency metrology to enable fit-for-purpose, cost-effective measurement solutions. Development of an ultra-stable temperature sensor that shows minimal drift over decadal time spans or a field-deployable thermodynamic temperature sensor, likely based on quantum technologies could disrupt the calibration-centered metrology ecosystem of today \cite{review2022, xu2014ultra}. Chip-based photonic and quantum thermometry technologies, such as photonic ring resonators, are well-suited for macroscale sensors that address the industrial needs of today. However, the emerging field of nanoscale heat transport in complex systems, such as quantum information systems \cite{review2022}, biological matrices, and advanced computer chips, demands the development of novel nanoscale temperature sensors.

In this context,  nitrogen-vacancy (NV) in diamond has emerged as a potential candidate technology to enable temperature measurements at micro and nanoscale distances. These sensors leverage the sensitivity of NV spin systems to environmental magnetic and electrical fields  \cite{maze2008nanoscale, dolde2011electric} along with sensitivity to local temperature and pressure to enable high sensitivity. While NV has garnered considerable attention for quantum magnetometry applications, its sensitivity to temperature  has led to suggestion that it may be suitable for temperature measurement applications in embedded systems such as temperature control in microfluidics \cite{neumann2013high}, generating heatmaps of the surrounding environment of thin metallic wires \cite{Toyli2013}, and measuring temperature of living cells \cite{kucsko2013nanometre}. Temperature sensitivities on the order of $10 \ \mathrm{mK}/\sqrt{\mathrm{Hz}}$ have been reported  \cite{kucsko2013nanometre, zhang2021robust} using pulsed optically detected magnetic resonance (pulsed-ODMR) highlighting the potential of NV thermometry. Recent technological advancements, such as integration with optical fiber probes \cite{zhang2021robust,fedotov2014fiber} have expanded the potential for practical applications of NV sensors. Using continuous-wave ODMR measurement, researcher's have reported sensitivities of $ 2 \ \mathrm{K}/\sqrt{\mathrm{Hz}}$\cite{Fujiwara_2021}.

In this work, we assess how different inference models affect uncertainties by comparing data-driven and model-based approaches for temperature estimation. The model-based approach relies on a probabilistic feedforward inference model based on NV's Hamiltonian that is used to infer temperature from NV's optically detected magnetic resonance (ODMR) spectra acquired using a fiber-coupled NV temperature sensor. This probabilistic model is compared to non-parametric peak-finding routines and unsupervised learning techniques such as Principal Component Regression (PCR) \cite{massy1965principal} and Convolutional Neural Network (CNN) \cite{o2015introduction}.  The later two methodologies are entirely data-driven and do not rely on any expert/physics-based insights on the relationship between spectroscopic features and temperature. Our results demonstrate physics-based probabilistic model is competitive with data-driven models. Data-driven models, when tested with out-of-sample data, outperforms the probabilistic model. However, when tasked with extrapolating beyond the training range,  data-driven models vastly underperform against the probabilistic model. 

\section{Methods}

Details of the experimental design (including sensor fabrication and instrumentation setup) can be found in the supplementary material; in this section we focus on data processing methods.

\subsection{Data processing and evaluation}
Spectral normalization ensures consistency across spectra by zero-centering and scaling the data. While some models are robust to unnormalized data, neural networks are particularly sensitive to amplitude variations near resonance frequencies. To ensure fair comparisons, a uniform normalization method is applied, effectively introducing a single additional parameter per model with minimal risk of overfitting. However, this process may also remove temperature-related information, suggesting that an ideal model leveraging unnormalized data should perform at least as well if it effectively captures amplitude variations.

Each temperature estimation method is evaluated using both in-sample and out-of-sample data. The in-sample data consists of measurements within the training range, while the out-of-sample data includes measurements from a different range or dataset not seen during training. This division allows for an assessment of each methodology’s ability to generalize across varying experimental conditions and sample variations. The predicted temperatures are compared against the measured values to evaluate accuracy, robustness, and potential for fine-tuning. The out-of-sample and in-sample prediction uncertainties for each model are presented in Table \ref{tab: out sample} and Table \ref{tab:in sample}, respectively. Additionally, in Table \ref{tab:residuals errors}, each model's output is augmented with a linear regression to correct for bias introduced by the linear assumption.

\begin{figure}[htbp]
\centering
     \begin{subfigure}[b]{0.489\textwidth}
         \centering
         \includegraphics[width=\linewidth]{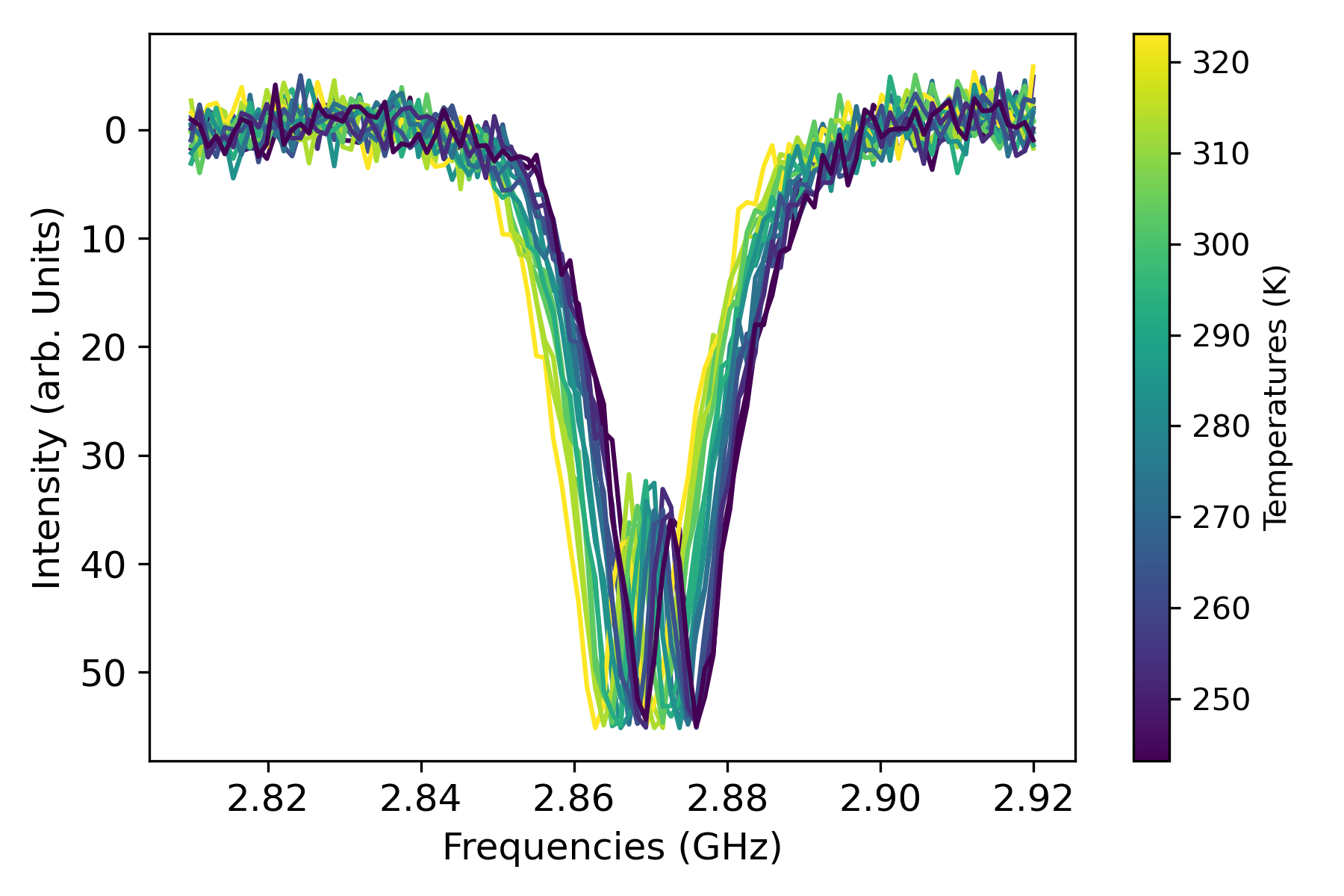}
         \label{fig:spec_curve}
     \end{subfigure}
       \hfill
     \begin{subfigure}[b]{0.489\textwidth}
         \centering
         \includegraphics[width=\linewidth]{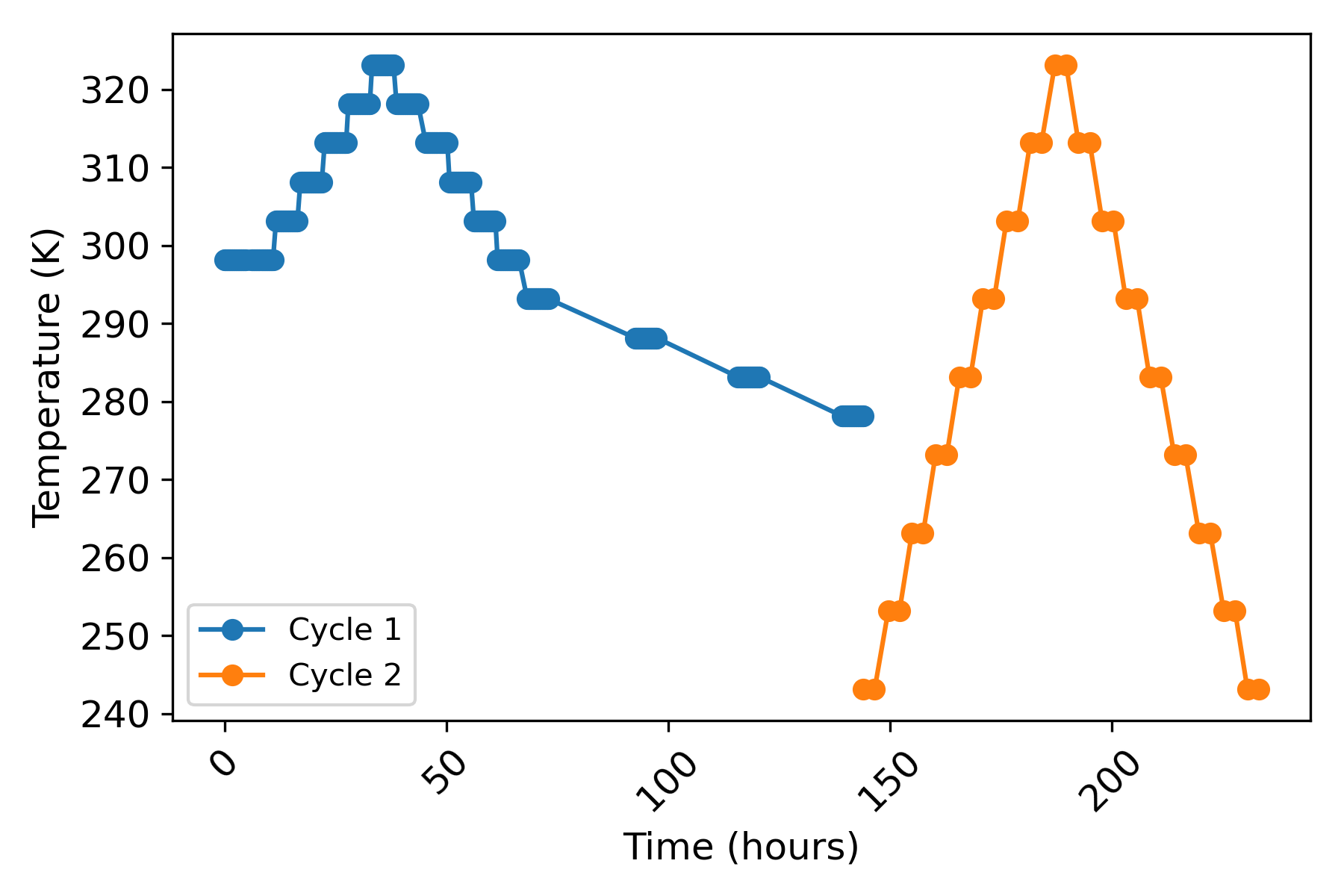}
         \label{fig:ind_poisson}
     \end{subfigure}
   \caption{Left: ODMR spectra across a temperature range of 243.14 K to 323.15 K. Note that the spectra do not show significant changes in contrast and width. Right: Temperature history of the sensor over cycle 1 and 2.}
    \label{fig: average spectrum}
\end{figure}

\subsection{ Estimating temperature: a model based approach}
\subsubsection{ Overview of the model}
The behavior of the NV center in diamond is governed by its spin Hamiltonian \eqref{eq:hamiltonian}, which describes the energy levels and dynamics of the electron spin\cite{doherty2013nitrogen}.

\begin{equation}
\hat{H} = \hbar D\left[\hat{S}_Z^2-\frac{2}{3}\right] + \hbar E\left(\hat{S}_X^2-\hat{S}_Y^2\right)
\label{eq:hamiltonian}
\end{equation}
where $\hbar$ is the reduced Planck constant, $D$ is the axial zero-field splitting parameter, $E$ is the non-axial zero-field splitting parameter, and $\hat{S}_X$, $\hat{S}_Y$, and $\hat{S}_Z$ are the spin operators. The axial parameter $D$ is particularly sensitive to changes in temperature and primarily governs the center frequency of the ODMR spectrum. In contrast, the non-axial parameter $E$ has a weaker temperature dependence and mainly influences the splitting frequency between resonance peaks in the ODMR spectrum. The temperature dependence of $D$ and $E$ is typically described by linear equations, as shown in equations \eqref{eq:D} and \eqref{eq:E}, where $T$ is the temperature and $a$, $b$, $c$, and $d$ are constants specific to the NV center system. Notably, constants \( c \) and \( d \) are treated as fixed using literature values \cite{rembold2020introduction} \( c = -4.65 \times 10^{-7} \, \text{GHz}/\text{K} \) and \( d = 0.0052 \, \text{GHz}/\text{K} \) \cite{range}.

\begin{equation}
D = aT + b
\label{eq:D}
\end{equation}
\begin{equation}
E = cT + d
\label{eq:E}
\end{equation}

The actual observed ODMR spectrum as shown in 
 \textcolor{black}{ Figure~\ref{fig: average spectrum}} is modeled as a combination of two Lorentzian functions (bi-Lorentzian), as shown in Figure~\ref{fig:bi-lorenz-fit}, centered at transition frequencies corresponding to the spin Hamiltonian's eigenvalues, with additional Gaussian noise. By incorporating temperature into the $D$ and $E$ parameters, which in turn affect the Hamiltonian's eigenvalues, it becomes possible to determine the peak centers in the ODMR spectrum. To fully specify the model, the widths and amplitudes of the Lorentzian functions, as well as the parameters $a$ and $b$ relating $D$ and $E$ to temperature, needs to be determined. Once these parameters are known, the model can calculate the likelihood of observing a given ODMR spectrum for different temperature values.

The key idea is to use a probabilistic feedforward inference approach \cite{griffiths2008primer} to find the temperature value that maximizes the likelihood of the observed ODMR spectrum, effectively utilizing the temperature dependence of the spin Hamiltonian parameters to infer the temperature from the observed spectral features in the ODMR spectrum.The use of Maximum Likelihood Estimation (MLE) provides an additional level of flexibility, namely, the model can be readily adapted to single spin-single photon measurements by changing the noise model from Gaussian to Poisson. In the case of ensemble measurements, a Gaussian noise model is appropriate and the MLE model behaves similar to a least squares model but also allows maximum likelihood parameter estimation.

 \subsubsection{Fitting model parameters}

\begin{figure}[htbp]
\centering
     \begin{subfigure}[b]{0.49\textwidth}
         \centering
         \includegraphics[width=\linewidth]
         {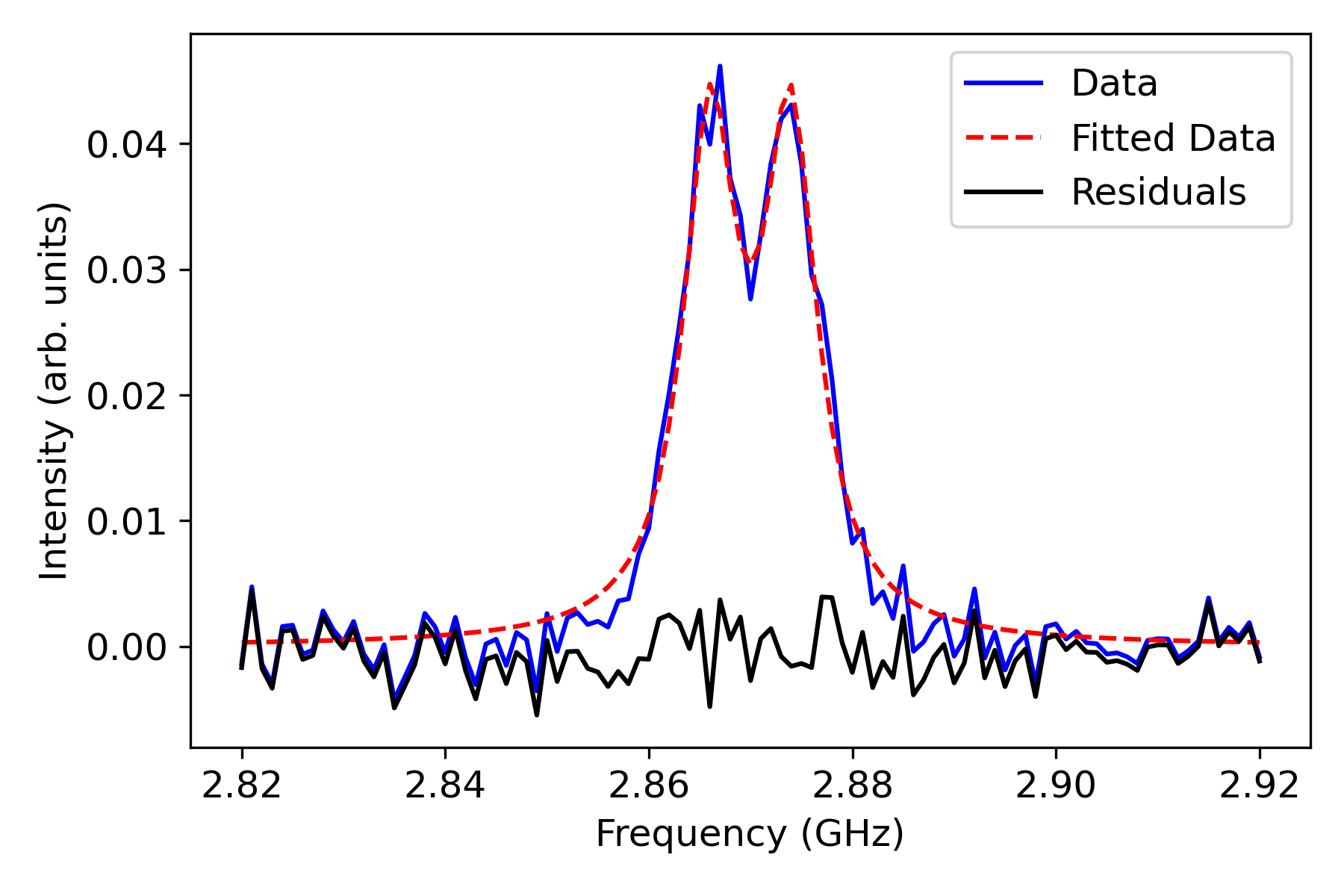}
        % \caption{Comparison of Measured and Predicted Temperatures }
         \label{fig:spec_curve}
     \end{subfigure}
       \hfill
     \begin{subfigure}[b]{0.49\textwidth}
         \centering
         \includegraphics[width=\textwidth]{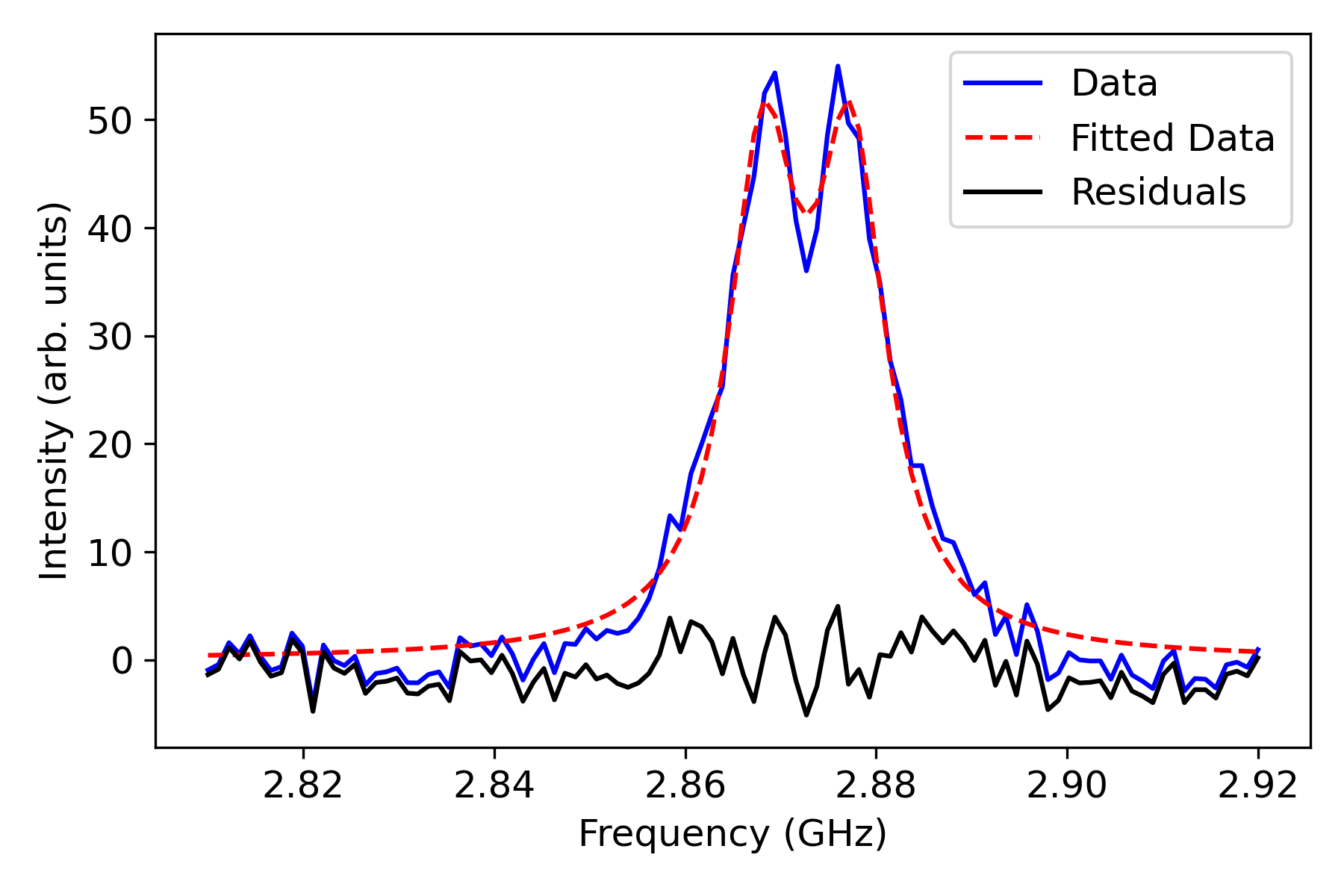}
         % \caption{Temperature Dependence of Zero-Field Splitting Parameter}
         \label{fig:ind_poisson}
     \end{subfigure}
    \caption{ ODMR spectra of NV chip (Element six)  and fiber coupled microparticle are fitted to bi-lorentzian function. The suitability of fit shows that 2 -Lorentzian function can be used across wide range of NV sensors regardless of their origin. The flat residuals indicate Lorentzian function is sufficient to capture spectral line shape. }
    \label{fig:bi-lorenz-fit}
\end{figure}

To fully specify the model, we conduct a fitting procedure using a bi-Lorentzian function applied to the observed ODMR spectrum.  As illustrated in Figure \ref{fig:bi-lorenz-fit}, bi-Lorentzians fit the ODMR profile of both a packaged i.e. fiberized NV sensor and an "unpackaged" single crystal chip suggesting the use of Lorentzians under low laser and microwave power is broadly applicable to bulk NV sensors. This fitting process allow us to determine the nominal values of amplitudes and widths of the Lorentzian components over the temperature examined here. Over the temperature examined here, amplitude and width show little temperature dependent variation, as such we treat these variables as being temperature independent in our model \cite{wider}.

% \footnote{We note, that over wider temperature ranges it is known that that amplitude (or contrast) shows a modest inverse relationship whilst width shows a modest positive correlation. Incorporating this relationship into the model is the subject of future work.} 

Once the Lorentzian parameters have been determined, we can proceed to estimate the temperature-dependent coefficients, $a$ and $b$, using Maximum Likelihood Estimation (MLE) optimization. The MLE approach involves constructing a likelihood function that quantifies the probability of observing the given spectroscopic data for the coefficients $a$ and $b$.

The overall likelihood for all spectra can be expressed as:

\begin{equation}\label{eq:likelihood_all}
L(\text{all spectrum} \mid a, b) = \prod_{i=1}^{M} L_i(\text{spectrum}_i \mid a, b)
\end{equation}

where \(M\) is the total number of spectra, and \(L_i(\text{spectrum}_i \mid a, b)\) is the likelihood of observing the \(i\)-th spectrum given \(a\) and \(b\). This product represents the assumption that the spectra are independent observations.

During model fitting, we sum the log-likelihoods of all spectra (equation \ref{eq:likelihood_all}) and maximize them with respect to \(a\) and \(b\) using Powell's conjugate direction method \cite{powell1964efficient}. This derivative-free optimization method possesses the capability to handle multiple variables and is applicable to non-linear functions. Its simplicity of implementation, which circumvents extensive parameter tuning, renders Powell's conjugate direction method a fitting choice for maximizing the log-likelihood function. Once $a$ and $b$ are determined, we fix these values and maximize the log-likelihood with respect to temperature for any single spectrum, as discussed in the subsequent section.

% \begin{equation}\label{eq:loglikelihoodab}
%   \log L(a, b \, | \, \text{spectroscopy data}) = \sum_{k=1}^N -\frac{(m_k - \Delta t r \mathcal{S}_k(a, b))^2}{2\sigma_k^2} - \log(\sqrt{2\pi}\sigma_k)   
% \end{equation} 
% % 

\subsection{ Running the model}

 Our objective is to determine the most accurate estimates for temperature using MLE based on spectroscopy data. This entails creating a likelihood function, which relies on Gaussian  distributions, and identifying the parameter values that maximize it.  The likelihood function assesses how well the bi-Lorentzian function \(\mathcal{S}_{k,i}(T)\) evaluated at the \( k \)-th frequency point for the \( i \)-th spectrum, with temperature $T$ fits the observed data  \( m_{k,i} \)  at a specific frequency $k$. It originates from photon observations, considering the probability of encountering a particular number of photons while factoring in the expected rate of photon occurrences over time.\\

Each measurement \( m_{k,i} \) is assumed to follow a Gaussian distribution centered around the expected value \( \mu_{k,i} \) with standard deviation \( \sigma_{k,i} \). The likelihood function for observing \( m_{k,i} \) given \( \mu_{k,i} \) is:

\begin{equation}
L(m_{k,i} \mid \mu_{k,i}) = \frac{1}{\sqrt{2\pi}\sigma_{k,i}} \exp\left(-\frac{(m_{k,i} - \mu_{k,i})^2}{2\sigma_{k,i}^2}\right)
\end{equation}

The expected value \( \mu_{k,i} \) is modeled by the bi-Lorentzian spectrum function, scaled by the known parameters \( \Delta t \) and \( r \)

\begin{equation}
\mu_{k,i} = \Delta t \, r \, \mathcal{S}_{k,i}( T)
\end{equation}

The likelihood of a single spectrum consisting of \( N_i \) measurements is given by:

\begin{equation}
L(m_{1,i}, m_{2,i}, \ldots, m_{N_i} \mid  T) = \prod_{k=1}^{N_i} \frac{1}{\sqrt{2\pi}\sigma_{k,i}} \exp\left(-\frac{(m_{k,i} - \Delta t \, r \, \mathcal{S}_{k,i}(a, b, T))^2}{2\sigma_{k,i}^2}\right)
\end{equation}

To facilitate optimization, we consider the log-likelihood function. Taking the natural logarithm of the likelihood function:

\begin{equation}\label{eq:loglikelihoodwhole}
\log \mathcal{L}_i(\text{spectrum}_i \mid T) = \sum_{k=1}^{N_i} \left[ -\log (\sqrt{2\pi}\sigma_{k,i}) - \frac{(m_{k,i} - \Delta t \, r \, \mathcal{S}_{k,i}(a, b, T))^2}{2\sigma_{k,i}^2} \right]
\end{equation}

Now, the focus is on optimizing the log-likelihood function (equation ~\ref{eq:loglikelihoodwhole}), which relies on the Hamiltonian via its eigenvalues. A significant hurdle in MLE arises when computing derivatives with respect to temperature. While theoretical formulas for these derivatives exist, as demonstrated in \cite{brewer1978kronecker}, modern programming favors more efficient methods for computation. Here we employ automatic differentiation as the preferred approach \cite{strang2019linear}.

To leverage automatic differentiation effectively, we rely on TensorFlow \cite{goldsborough2016tour}, a widely adopted library for machine learning and scientific computation. In order to quickly maximize the log-likelihood function, and find the MLE temperature, it is helpful to find the derivative of the log-likelihood with respect to temperature, however this is challenging since computing the log-likelihood involves solving an eigenvalue problem for the Hamiltonian.  TensorFlow enables us to easily compute the this complicated derivative, by using automatic differentiation to find the derivative of the eigensolver. Subsequently, we utilize the optimization functionalities provided by the SciPy \cite{virtanen2020scipy} library. Specifically, we employ the Newton-CG (Conjugate Gradient) optimization algorithm, which is a specialized refinement of Newton’s method tailored for problems with small feature spaces. Newton's method typically involves computing and inverting the Hessian matrix, which can be computationally intensive for large feature sets. However, Newton-CG is designed to address this issue by combining Newton’s method with the Conjugate Gradient approach, thereby avoiding the direct computation of the Hessian inverse. This hybrid approach facilitates rapid convergence while managing computational efficiency, making it well-suited for scenarios with a relatively small number of features. The algorithm iteratively refines our parameter estimates by minimizing the negative log-likelihood function until convergence is achieved.

\begin{figure}[htbp]
\centering
\begin{subfigure}[b]{0.49\textwidth}
    \centering
    \includegraphics[width=\linewidth]{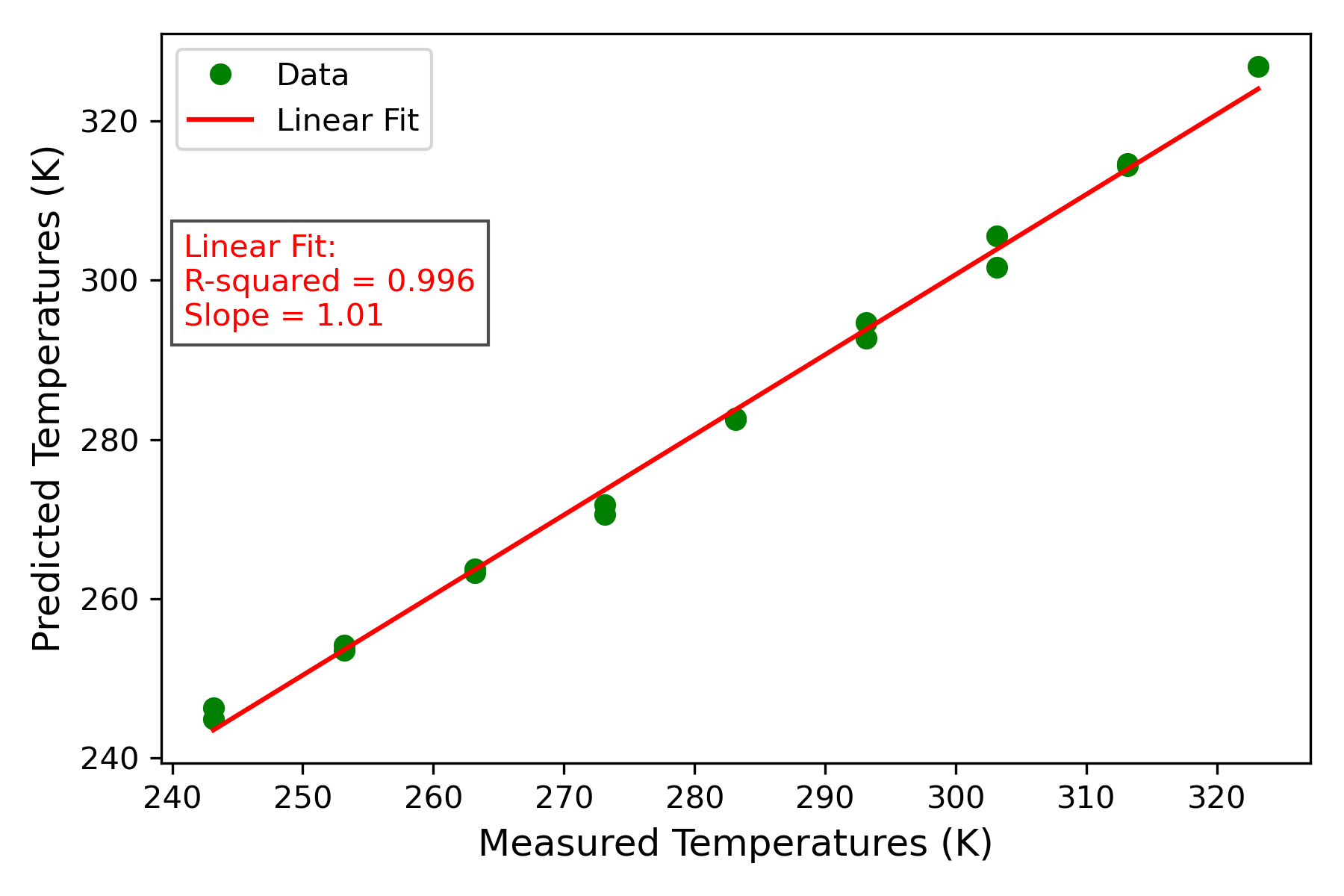}

\end{subfigure}
\hfill
\begin{subfigure}[b]{0.49\textwidth}
    \centering
    \includegraphics[width=\linewidth]{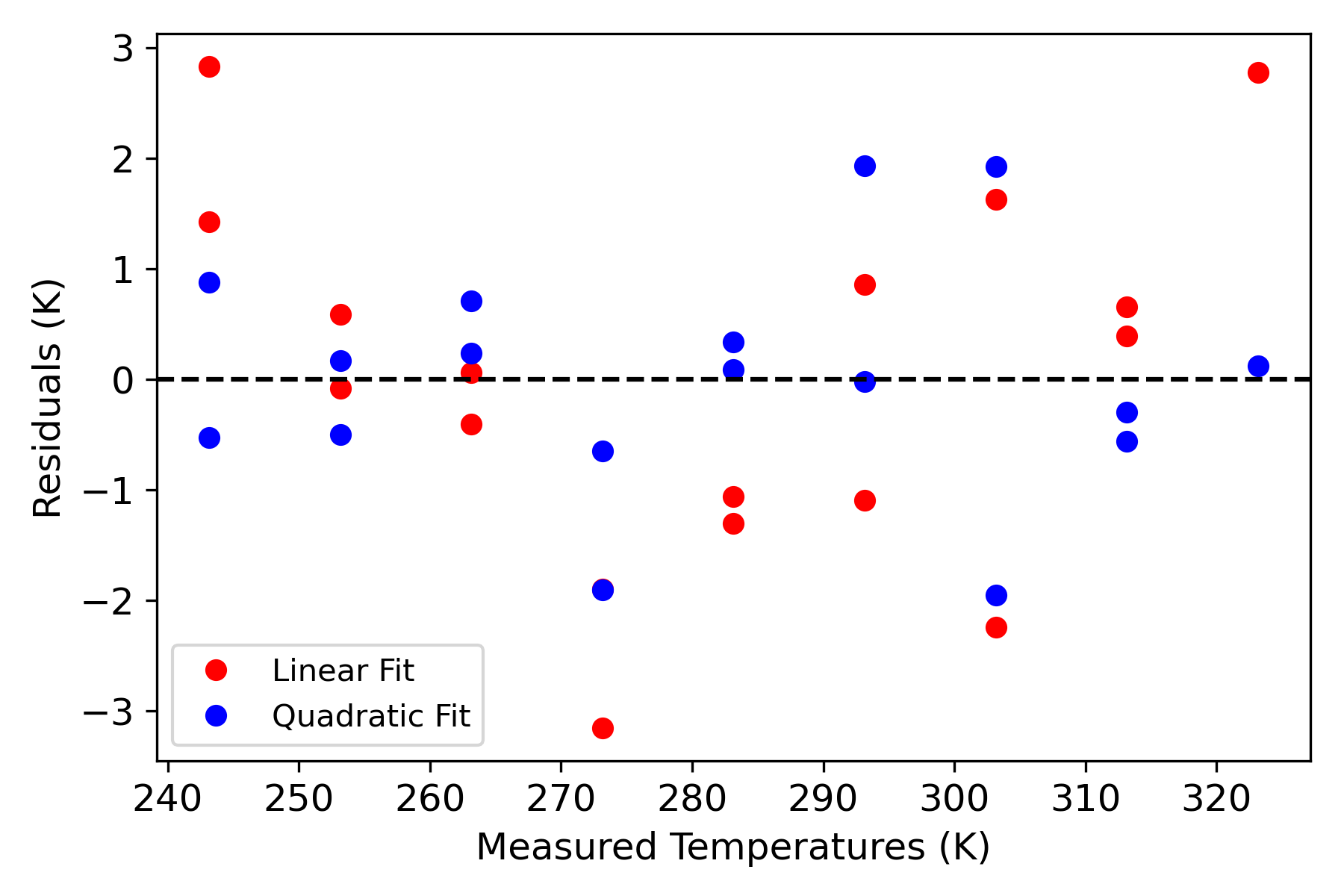}
    
\end{subfigure}
\caption{ Left: Cycle 2 (243.14 K to 323.15 K) data fit with parameters from Cycle 1 (283.5 K to 323.15 K), showcasing cross-cycle calibration capability. Right: Residuals shown against temperature.}

\label{fig:cycle2fit}
\end{figure}

% Explain about cycles of sensor 1
The depiction in Figure ~\ref{fig:cycle2fit} sheds light on the excellent correlation between temperature reported by the drywell and predicted temperatures during cycle 2 of a fiber-coupled NV sensor using parameters derived from cycle 1. The close match between the drywell's measured temperature and the model's predicted temperatures demonstrates the effectiveness of our method in capturing temperature variations within spectroscopy data. This is evidenced by a root mean squared error (RMSE) of $< 2$ K, as shown in Table \ref{tab: out sample} and \ref{tab:in sample}.

 We note that fitting cycle 2 using parameters specific to cycle 2 obtained nearly identical results with an RMSE between measured and predicted temperatures of $1.74 $ K, indicating little or no hysteresis between the two thermal cycles. The residuals associated with cycle 2 shown in Figure ~\ref{fig:cycle2fit} (right panel) are zero-mean centered and homoscedastic. We observe that the residuals for cycle 2, when fitted with quadratic functions, have a smaller magnitude of $1.02$ K compared to $1.62$ K with a linear fit. This suggests that the linear assumption introduces a slight bias into the model. Once corrected for bias, the  prediction uncertainties fall to 1.03 K (Table \ref{tab:residuals errors}).

\subsection{Estimating temperature using  Automatic  Peak Detection (AP)}
\subsubsection{Overview of the model}

The probabilistic model described in the previous section relies heavily on expert knowledge to parameterize the relationship between observed spectra and temperature. It offers a holistic perspective of the entire curve, leveraging its shape and dimensions to accurately determine it's temperature dependence. In contrast, the auto peak finder method, merely leverages the expert knowledge to posit that changes in peak (or valley) positions are correlated to changes in temperature. 

The peak finding algorithm pinpoints the positions of significant peaks by closely scrutinizing the local characteristics of the data. It achieves this by examining the behavior of two neighboring data points surrounding a peak to precisely identify its location. While the probabilistic model provides a comprehensive understanding of the curve's structure, the peak finding algorithm prioritizes localized features, allowing for efficient (fast) and targeted peak detection within the dataset.
\subsubsection{Running the model}
We begin by using the Python library \href{https://peakutils.readthedocs.io/en/latest/reference.html}{peakutils} to identify the indices of the peaks. To facilitate the algorithm we invert the spectra such that valleys appear as peaks.The indices identified by the algorithm as corresponding to peak locations are then sorted according to the corresponding y values (amplitudes). For the highest peak, a window of data points around the peak index is selected. A quadratic function is then fitted to these data points using the method of least squares. The location of the maximum of the fitted quadratic function is taken as the estimated peak location. This process is repeated for the next highest peak, excluding the data points already used for the previous peak(s).  The estimated peak locations are stored for further analysis and used to generate training and testing datasets. A linear regression model is then constructed to relate the temperature ($T$) to the peak locations, $P_1$ or $P_2$:  %axial ($D$) and non-axial ($E$) zero-field splitting parameters:
\begin{equation}
T = \alpha_0 + \alpha_1 P %+ \alpha_2 E
\end{equation}
where $\alpha_0$ and $\alpha_1$  are the coefficients to be determined.

\begin{figure}[htbp]
\centering
\begin{subfigure}[b]{0.49\textwidth}
    \centering
    \includegraphics[width=\linewidth]{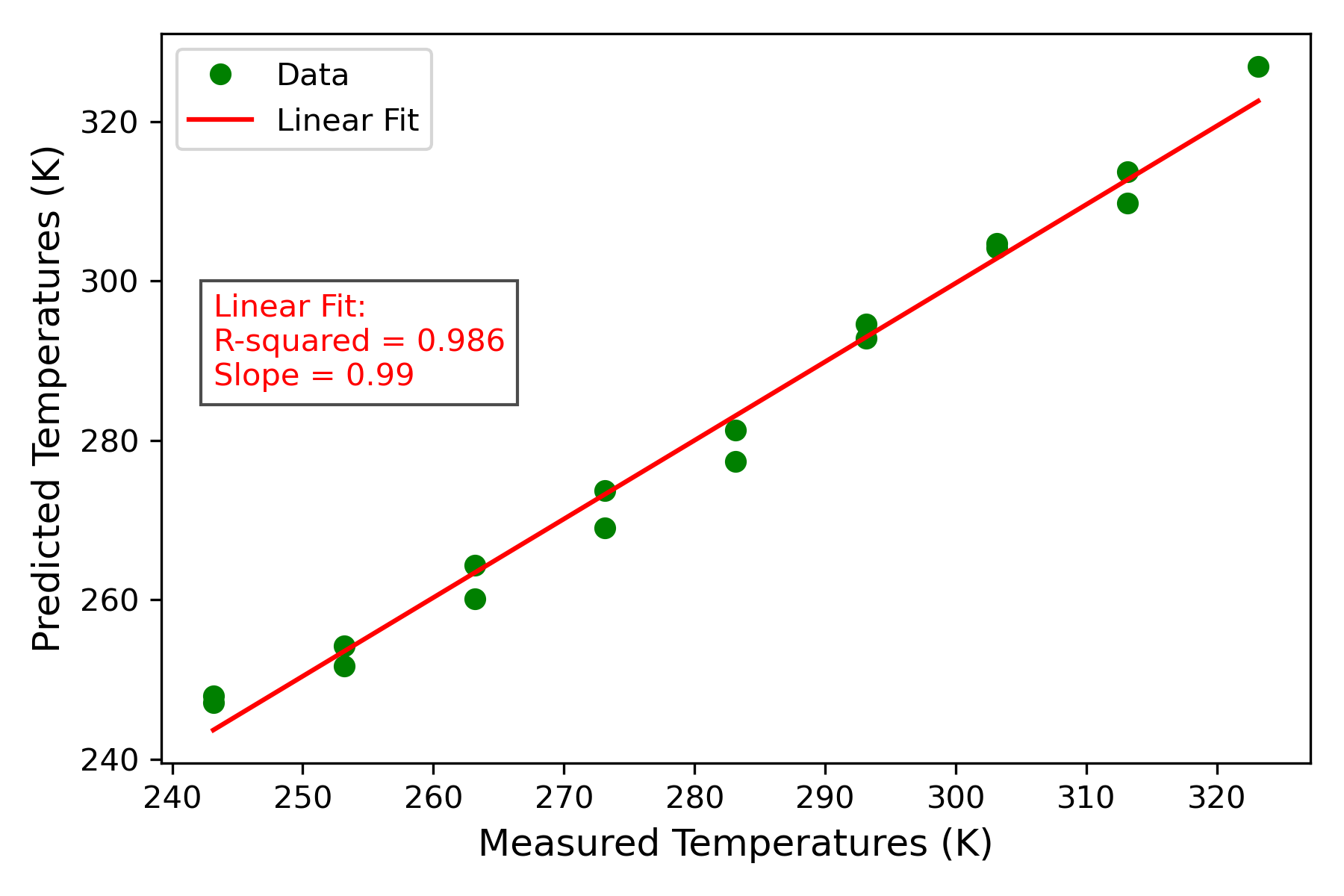}

\end{subfigure}
\hfill
\begin{subfigure}[b]{0.49\textwidth}
    \centering
    \includegraphics[width=\linewidth]{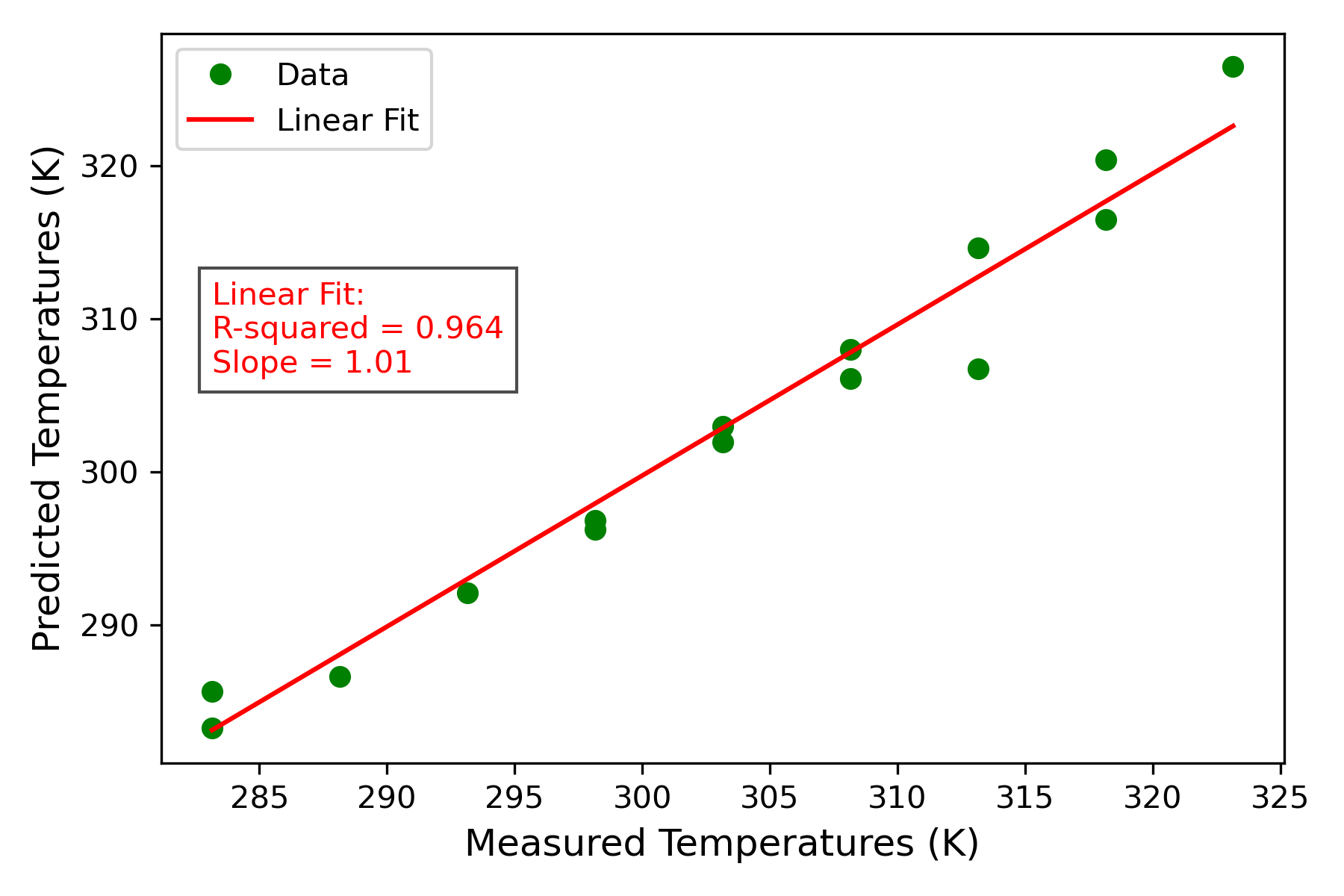}
    
\end{subfigure}
\caption{ Left : The model was trained on the peak locations extracted from the ODMR spectra of cycle 2 (243.14 K to 323.15 K). Right : Tested on the peak locations from cycle 1 (283.5 K to 323.15 K). The training and testing errors quantify the model's performance in predicting temperatures from the peak locations.}

\label{fig:Peaktrain2test1}
\end{figure}

Using the training dataset, which consists of the peak locations and the corresponding temperatures, the linear regression model is trained to determine the values of $\alpha_0$, $\alpha_1$, and $\alpha_2$. After training, the linear regression model predicts temperatures for the testing dataset, and its performance is assessed by comparing these predictions with the actual temperatures, resulting in the calculation of the testing error. Additionally, the model generates temperature predictions for the training dataset, allowing for the computation of the training error. Both the testing and training errors are depicted in Figure ~\ref{fig:Peaktrain2test1} and results summarized in Tables 1-3.  These results show the peak finder method results in prediction uncertainties that are upto 67$\% $ larger than than the probabilistic model. Post-bias correction the AP model's prediction uncertainty (2.30 K) is more than double the prediction uncertainty of probabilistic model.

\subsection{Estimating temperature using a data driven approach}
\subsubsection{Overview of the model}

The proposed methodology adopts a model-free approach known as Principal Component Regression (PCR) to infer temperature from ODMR spectra. The key steps of PCR are: (1) Principal Component Analysis (PCA) \cite{massy1965principal, beattie2021exploration} for dimensionality reduction and (2) linear regression for establishing the temperature-spectral relationship. This approach contrasts with the model-based approach in that it does not make make any assumptions about the distribution of spectral data nor it's temperature dependence. It relies entirely on the presented data to learn the relationship between spectra and temperature.

PCA serves as a robust tool for extracting the dominant modes of variation within the spectral dataset. By identifying these principal components (PCs), PCA facilitates dimensionality reduction while preserving essential information concerning temperature-dependent spectral features. Given the potentially high dimensionality of ODMR spectra, PCA is pivotal for capturing pertinent information within a lower-dimensional subspace.

In the PCA step, the modes (principal components) of the data are obtained by solving the eigenvalue problem:

\begin{equation}
    C U = U S
    \label{eq:4}
\end{equation}
where \( C \) represents the \( p \times p \) covariance matrix of the \( n \times p \) data matrix \( X \), \( S \) denotes the \( p \times p \) diagonal matrix of eigenvalues, and \( U \) is the \( p \times p \) matrix whose columns are the eigenvectors of \( C \). To reduce dimensionality, the top \( k \) eigenvectors corresponding to the largest eigenvalues are selected. For our purposes, \( k \) is set to 3, meaning that we choose the top 3 eigenvectors. These eigenvectors form the principal components that capture the most variance in the data.  Subsequently, the data are projected onto these principal components using the following equation:

\begin{equation}
    Y = (X - \mu) U_{k}
    \label{eq:5}
\end{equation}
In this equation, \( X \) represents the original data matrix, which is first centered by subtracting the mean vector \( \mu \). The matrix \( U_{k} \) is of size \( p \times k \) and contains the top \( k \) eigenvectors corresponding to the largest eigenvalues. %For this case, \( k = 3 \), so \( U_{k} \) consists of the top 3 eigenvectors. %These eigenvectors form the principal components that explain the most variance in the data.
The projection process involves multiplying the centered data matrix \( (X - \mu) \) by \( U_{k} \), resulting in a new data matrix \( Y \) with dimensions \( n \times k \). Here, \( n \) is the number of data points and \( k = 3 \) is the number of selected principal components. Each column of \( Y \) represents the coordinates of a data point in the 3-dimensional space spanned by the principal components.  Following dimensionality reduction via PCA, any form of regression can be employed to predict the temperature $T$ from the projected data $Y$. This is expressed as:

\begin{equation}
    T = f(Y,\theta)
\end{equation}
where $\theta$ represents the regression parameters. For instance, in linear regression, one fits $T = \varphi Y + \beta$, while in quadratic regression, one fits $T =  \psi y^2 + \varphi y + \beta $, where $y$ denotes the independent variable (i.e., the projected data $Y$). By learning a linear mapping between the reduced-dimensional spectral representation and the temperature values, the regression model provides a straightforward means of temperature inference.

\subsubsection{Running the model}

As noted above, the PCR model consists of first applying PCA as a feature identification step, and then applying a linear regression.  The method is first trained on a subset of the available data, referred to as the in-sample data. PCA is applied to this in-sample data  to identify the principal components. These principal components serve as a reduced-dimensional representation of the spectral features, retaining the essential information related to temperature dependence. Next, linear regression is performed on the projected in-sample data, where the principal component scores are used as input features, and the corresponding temperature values are used as the target variable. This regression learns the linear mapping between the reduced-dimensional spectral representation and the temperature values, effectively establishing the temperature-spectral relationship.

After training the model, it can be effectively applied to out-of-sample data, using the same mean \( \mu \) and modes \( U \) learned from the in-sample data, as described in equations \eqref{eq:4} and \eqref{eq:5}. This out-of-sample data comprises new ODMR spectra that were not included in the initial training set. These out-of-sample spectra undergo identical pre-processing steps, encompassing normalization and projection onto the previously learned principal components. Subsequently, the projected out-of-sample data are input into the trained linear regression model, which then predicts corresponding temperature values based on the learned temperature-spectral relationship. Notably, this prediction step is computationally efficient and can be performed in real-time, rendering the approach highly suitable for practical applications.

\begin{figure}[htbp]
\centering
\begin{subfigure}[b]{0.49\textwidth}
    \centering
    \includegraphics[width=\linewidth]{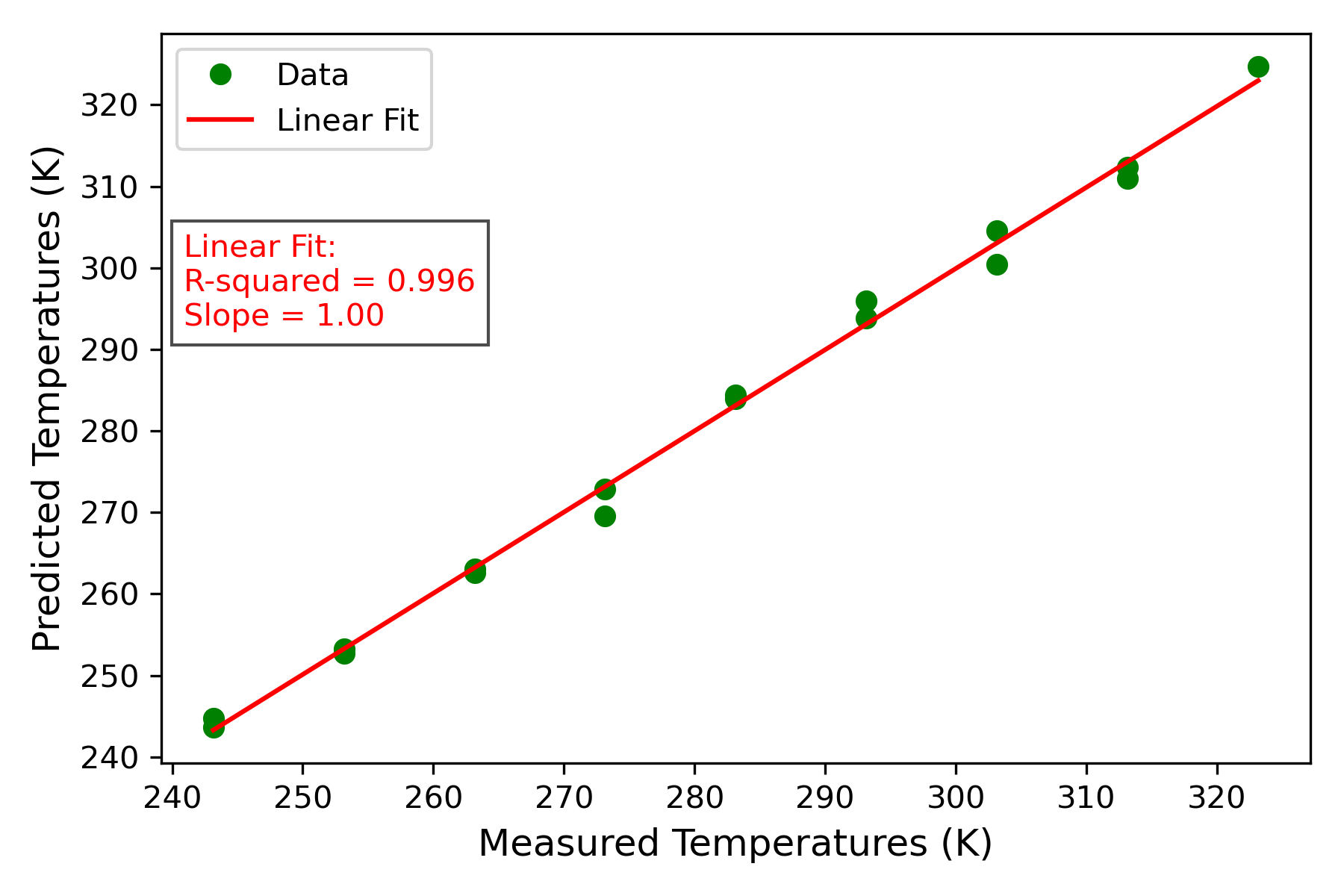}

\end{subfigure}
\hfill
\begin{subfigure}[b]{0.49\textwidth}
    \centering
    \includegraphics[width=\linewidth]{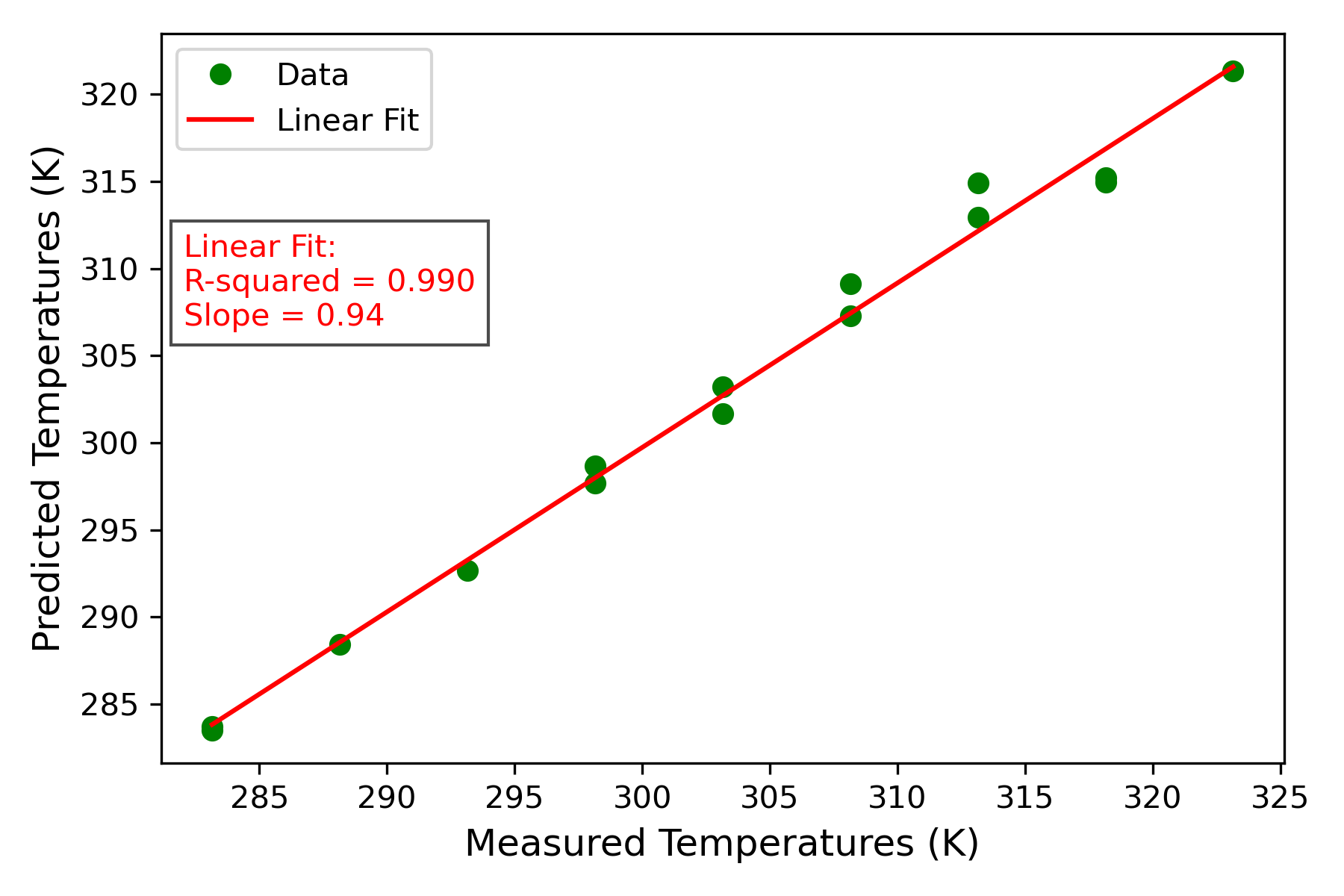}
    
\end{subfigure}
\caption{ Left: Training Error - Performance of the PCA regression model on cycle 2 (243.14 K to 323.15 K) of the sensor. Right: Testing Error - Evaluation of the PCA regression model's performance on cycle 1 (283.5 K to 323.15 K) of the same sensor.}

\label{fig:PCAtrain2test1}
\end{figure}

The proposed data-driven approach offers a robust and efficient solution for temperature inference from ODMR spectra. By integrating PCA for dimensionality reduction and linear regression for establishing the temperature-spectral relationship, this methodology provides a versatile framework adaptable to various experimental conditions, enabling accurate temperature prediction from ODMR spectra, even for out-of-sample scenarios. As shown in Fig ~\ref{fig:PCAtrain2test1} the data driven model readily learns the underlying mapping between temperature and eigenmodes of the PCA, with testing and training errors that are up to $0.67$ K lower than the statistical model \cite{regularization}.

\begin{figure}[htbp]
\centering
    \begin{subfigure}[b]{0.32\textwidth}
        \centering
        \includegraphics[width=\linewidth]{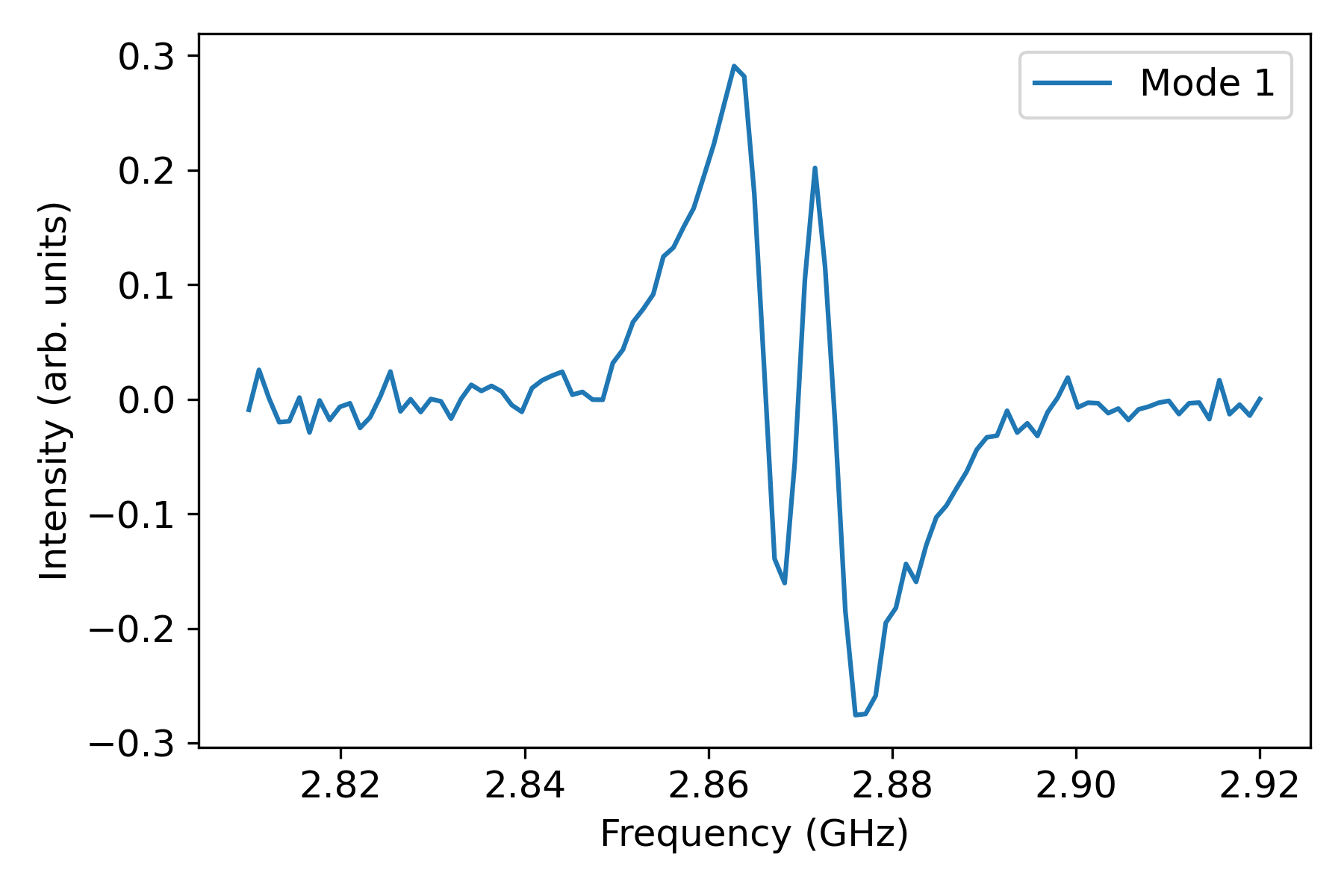}
    \end{subfigure}
    \hfill
    \begin{subfigure}[b]{0.32\textwidth}
        \centering
        \includegraphics[width=\linewidth]{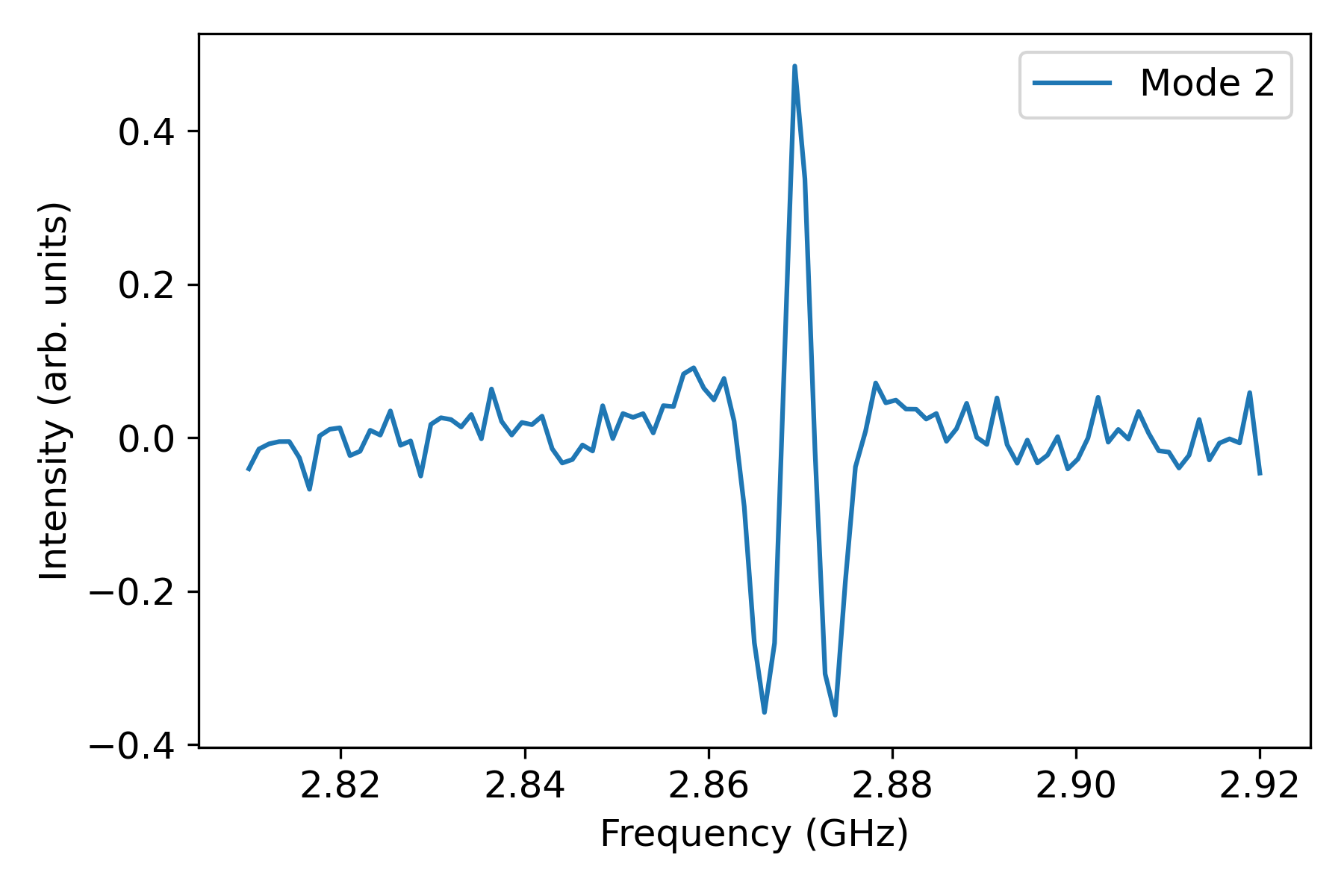}
    \end{subfigure}
    \hfill
    \begin{subfigure}[b]{0.32\textwidth}
        \centering
        \includegraphics[width=\linewidth]{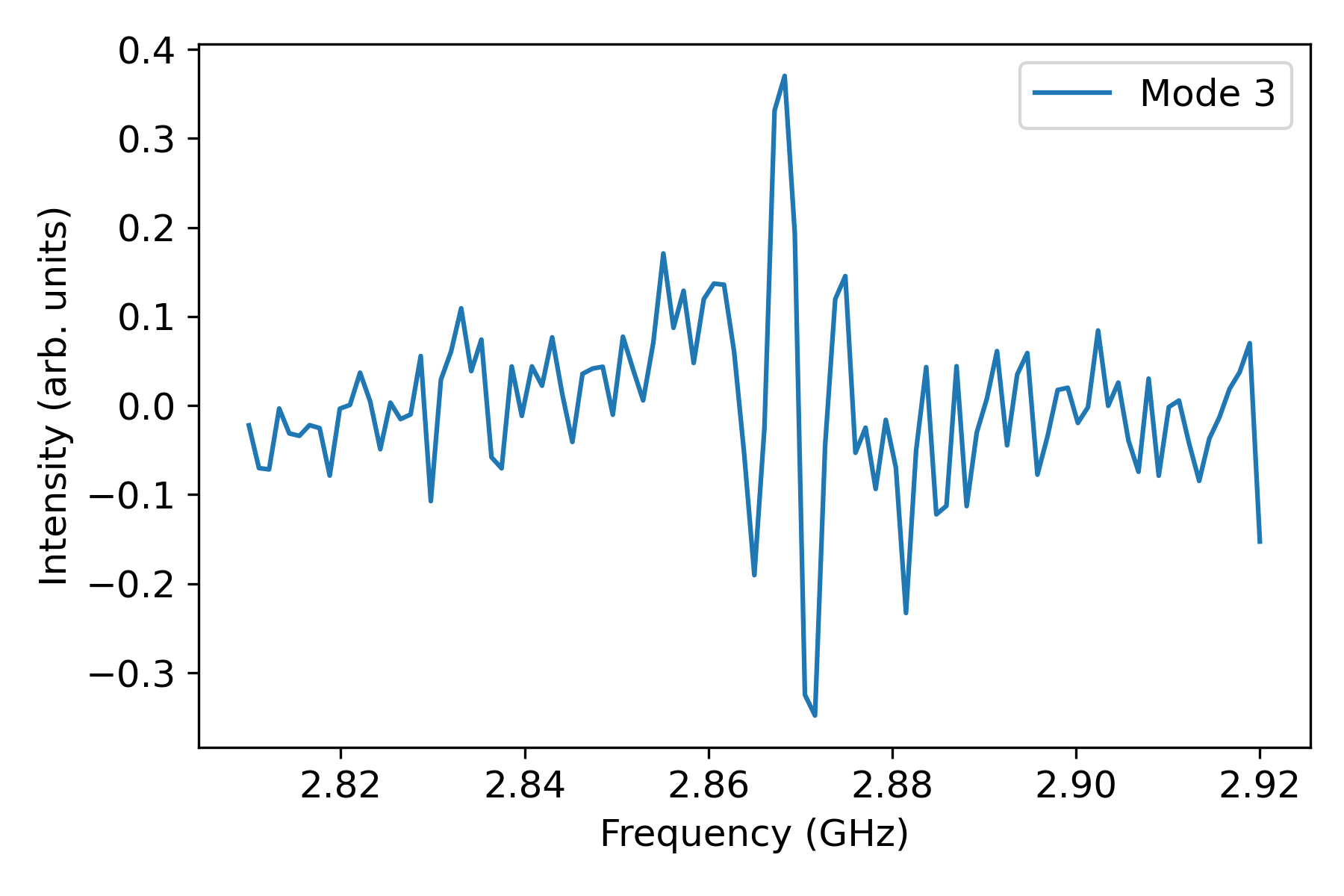}
    \end{subfigure}
    
    \begin{subfigure}[b]{0.32\textwidth}
        \centering
        \includegraphics[width=\linewidth]{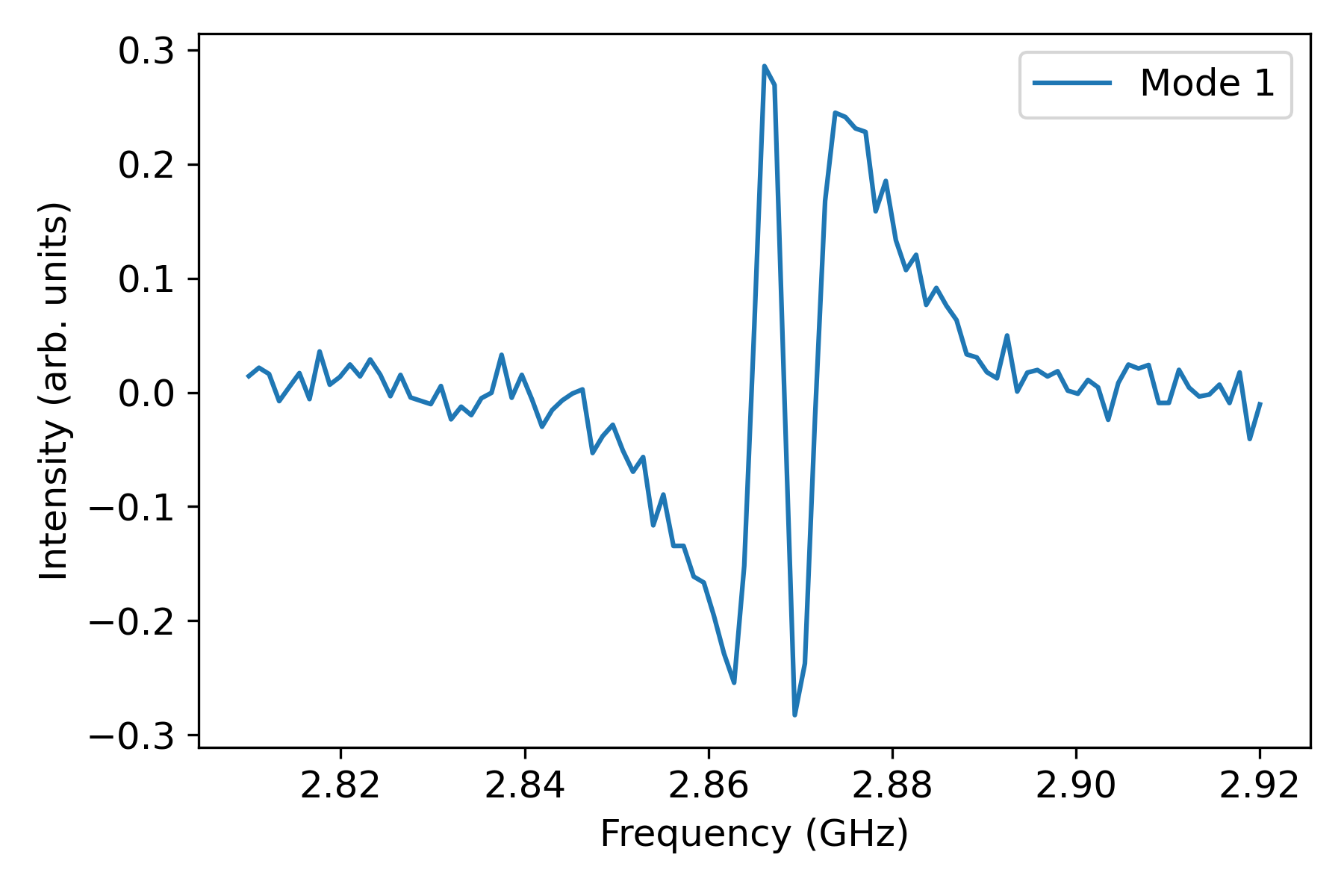}
    \end{subfigure}
    \hfill
    \begin{subfigure}[b]{0.32\textwidth}
        \centering
        \includegraphics[width=\linewidth]{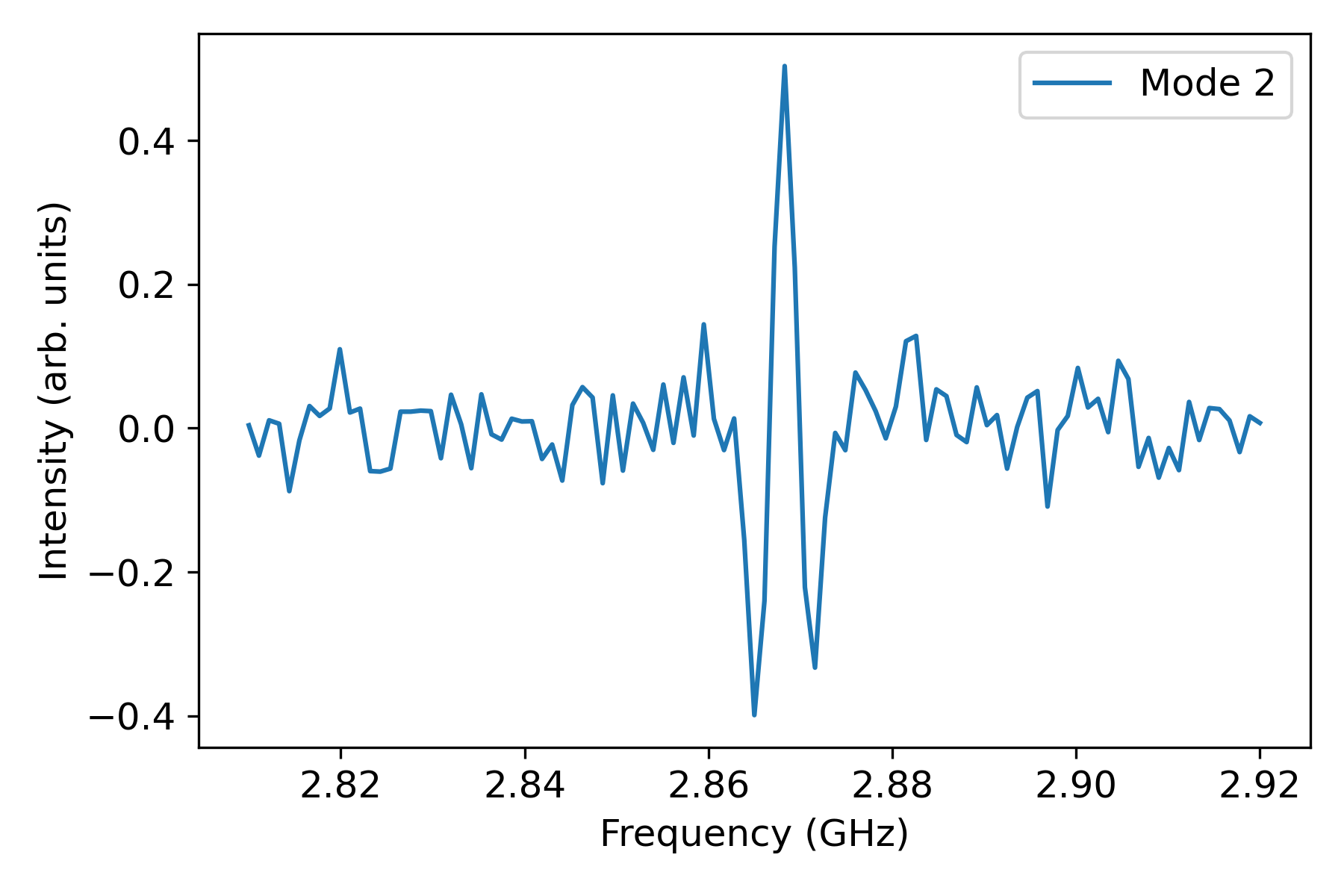}
    \end{subfigure}
    \hfill
    \begin{subfigure}[b]{0.32\textwidth}
        \centering
        \includegraphics[width=\linewidth]{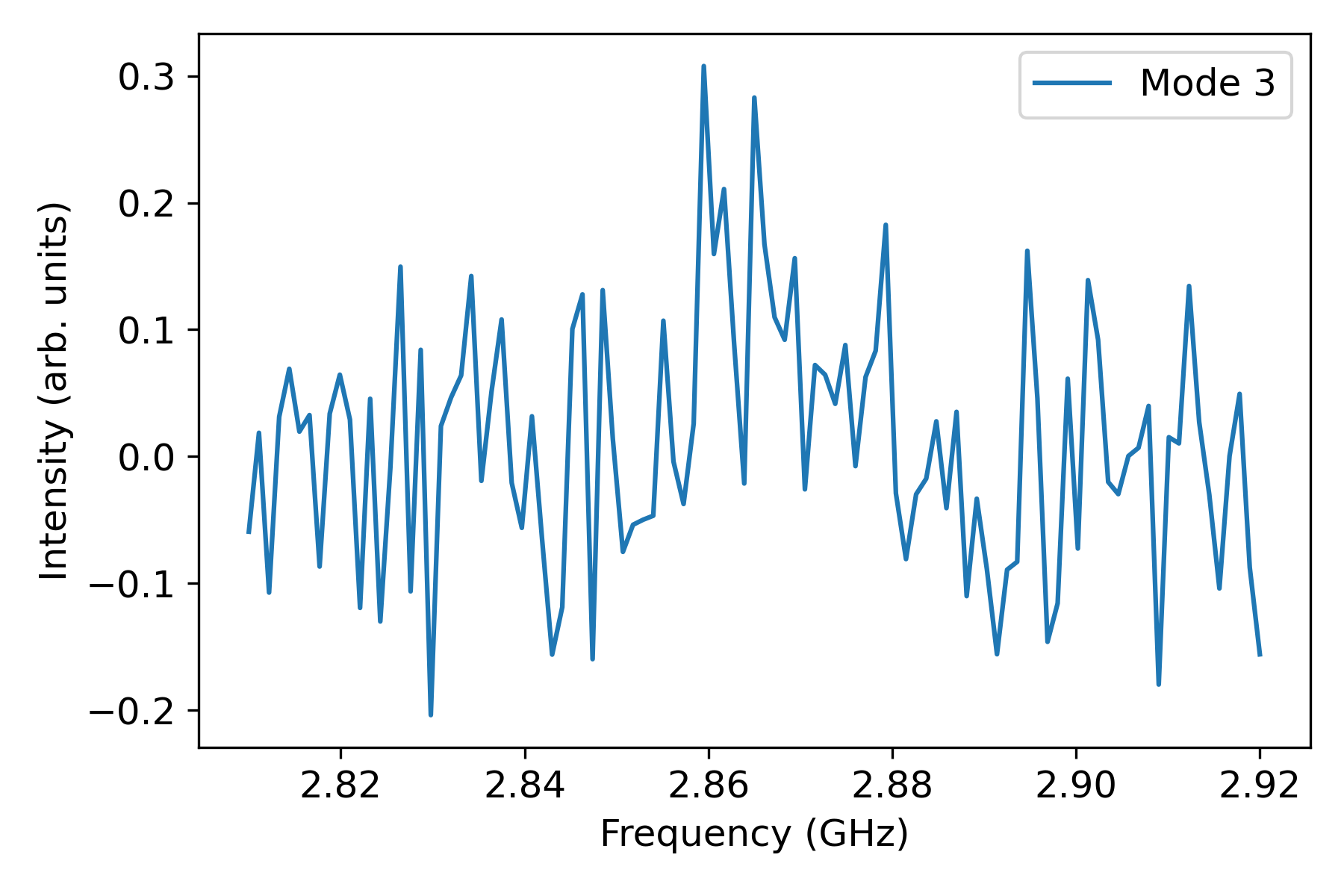}
    \end{subfigure}
    
    \caption{Modes obtained from Principal Component Analysis (PCA) regression modeling for two different temperature ranges. Top: Modes representing the dominant patterns of variation in the data for the temperature range of 243.14 K to 323.15 K. Bottom: Modes capturing the principal components of the data within the narrower temperature range of 283.5 K to 323.15 K. The modes are visualized as spatial patterns, revealing the characteristic features or structures associated with the respective temperature ranges.}
    \label{fig:Modes}
\end{figure}

\subsection{Estimating temperature using Convolutional Neural Network (CNN)}

\begin{figure}[htbp]
\centering
\begin{subfigure}[b]{0.49\textwidth}
    \centering
    \includegraphics[width=\linewidth]{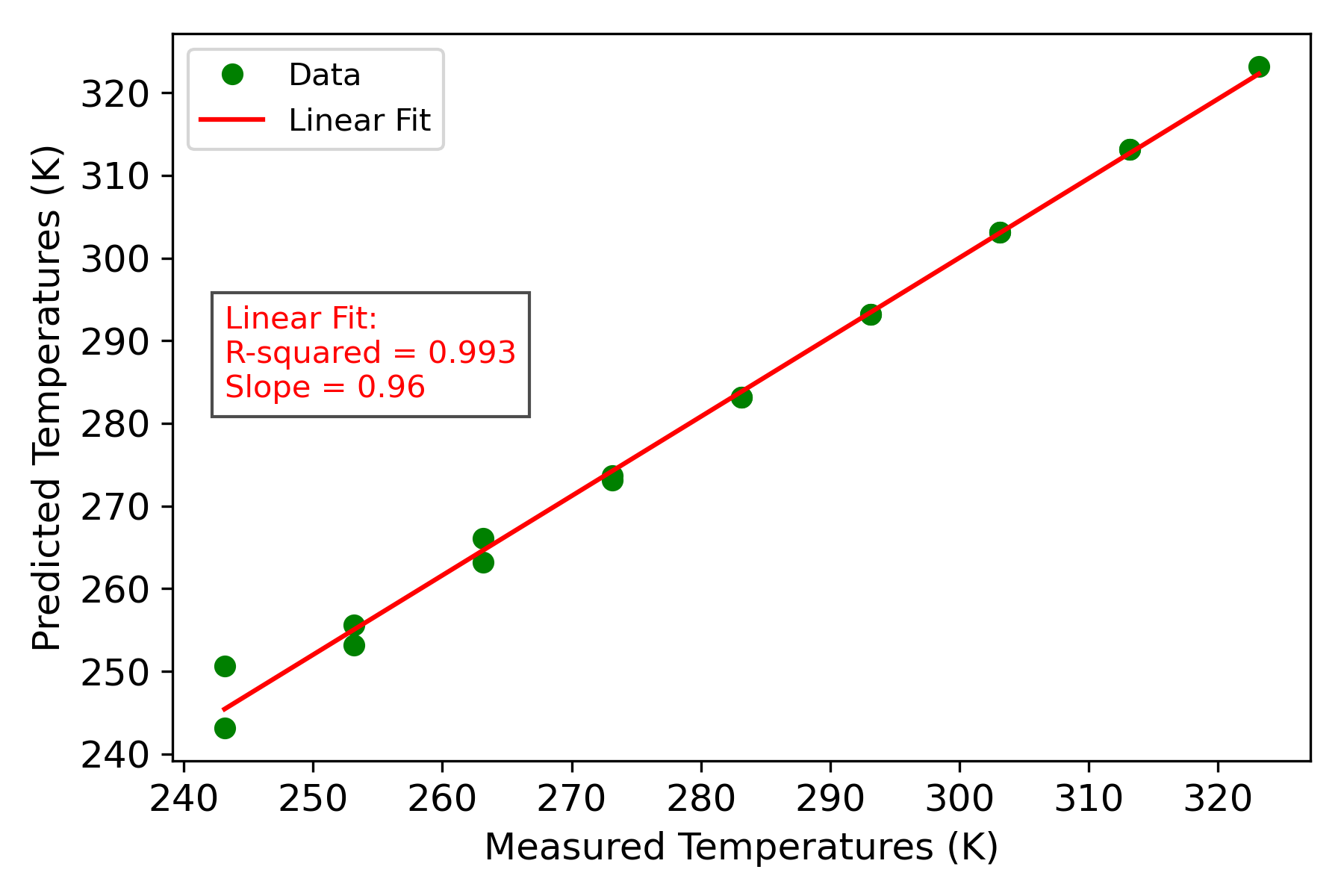}

\end{subfigure}
\hfill
\begin{subfigure}[b]{0.49\textwidth}
    \centering
    \includegraphics[width=\linewidth]{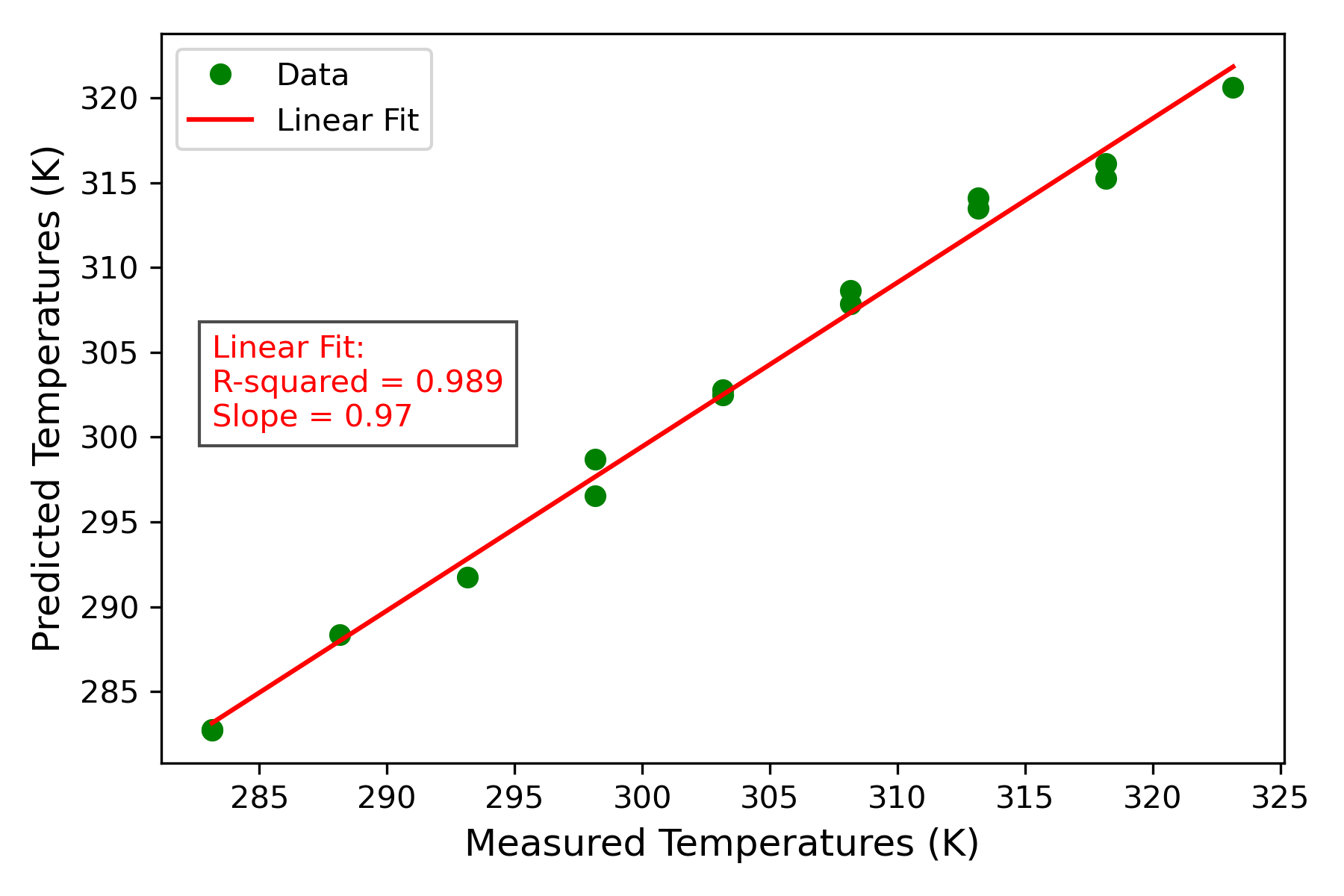}
    
\end{subfigure}
\caption{ Left: Training error of the CNN model, which was trained on the cycle 2 (243.14 K to 323.15 K) dataset from the sensor. Right: Testing error of the CNN model, which was tested on the cycle 1 (283.5 K to 323.15 K) dataset from the sensor. The plot depicts the discrepancy between the CNN's predicted temperatures and the actual temperatures. As the CNN iteratively optimizes its parameters, guided by the loss function, it progressively enhances its ability to capture the intricate temperature distributions even in the absence of labeled data.}

\label{fig:NNtrain2test1}
\end{figure}

\begin{figure}[htbp]
\centering
            \includegraphics[width=\linewidth]{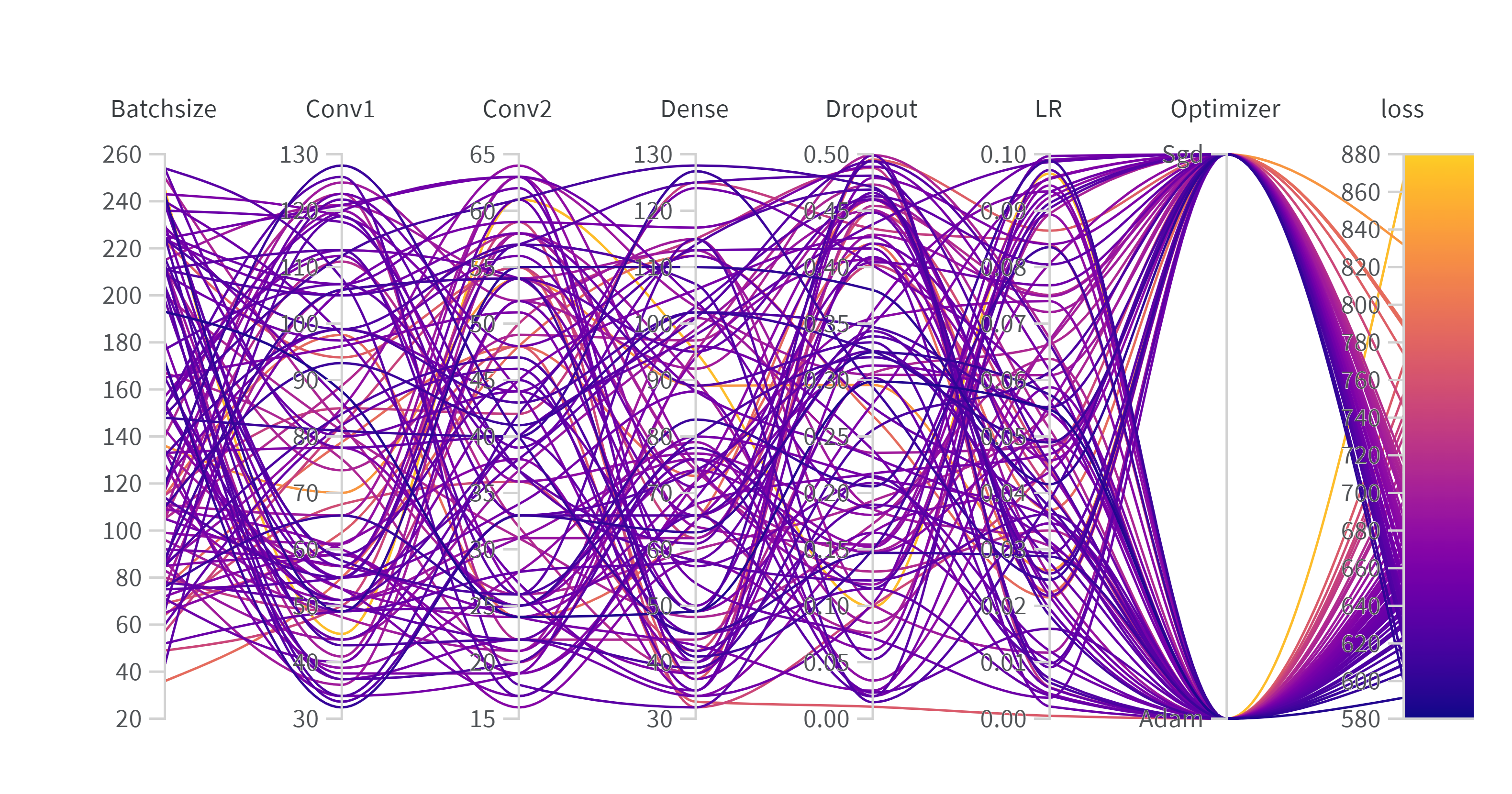}
        \hfill
    \caption{The figure is a visualization of the hyperparameter sweep performed over the sensor data. Curves that are closer in shape to dark purplish correspond to models that yield high performance, with respect to training loss, whereas yellow and orange are sub optimal. 
    Image created using Weights and Biases.\cite{guha2005interpreting}. Conv1 and Conv2 refer to the convolutional layers 1 and 2, respectively while LR represents learning rate. }
    \label{fig:Hyperparameter}
\end{figure}

We trained a multi-layer perceptron (MLP) using raw spectroscopic data. The MLP architecture included a flatten layer, two dense layers, and an output layer with ReLU activation functions. We conducted a hyperparameter sweep that systematically varied the hidden layer sizes, learning rates, and weight decay factors. Specifically, we used a learning rate of 0.001, with the first dense layer having 48 nodes, the second dense layer having 20 nodes, and the output layer set to 1. The MLP achieved a testing error of 3.14 K which is considerably larger than other models including the autopeak finder alogrithm. To improve upon MLP's performance we decided to use a more sophisticated approach: a 1D-Convolutional Neural Network. \textcolor{black}{Convolutional Neural Networks (CNN), in particular 2D-CNN are commonly used in computer vision applications, while 1D-CNNs have been applied to signal processing tasks. More recently CNNs have been applied to complex problems in materials science including inverse design \cite{jiang2021deep,li2022smart, li2021deep}.}
Here we employed the Sequential API from Keras as the foundation for developing a Convolutional Neural Network specifically tailored for temperature prediction. This model architecture, organized as a sequential stack of layers, is adept at handling sequential data such as the ODMR spectra. It encompasses Convolutional Layers for extracting features, MaxPooling1D Layers for dimension reduction, a Flatten Layer to prepare data for subsequent fully connected layers, Dense Layers for making predictions, and a Dropout Layer for regularization. Crucial hyperparameters, such as filter numbers, kernel size, dense layer units, dropout rate, and learning rate, are meticulously fine-tuned using Keras tuner \cite{omalley2019kerastuner}. The CNN model, compiled with the Adam optimizer \cite{kingma2014adam} and equipped with a loss function for measuring prediction accuracy, dynamically learns pertinent features from the input data, enabling precise temperature forecasts even for unseen data. Through extensive training on the dataset, the CNN model effectively discerns intricate patterns and relationships, culminating in accurate temperature predictions (Figure ~\ref{fig:NNtrain2test1}). This process is complemented by the visualization of the hyperparameter sweep all illustrated in Figure~\ref{fig:Hyperparameter}. This visual exploration allows for fine-tuning critical parameters, ensuring the CNN model attains its lowest validation loss and achieves optimal predictive capability.  \textcolor{black}{As shown in Fig \ref{fig:Hyperparameter} clustering observed in the visualization suggests that the search space is not flat; rather, there are distinct regions where certain hyperparameter combinations, particularly those involving Conv1, Conv2, and learning rate (LR), lead to better model performance. This structured distribution highlights the presence of optimal regions within the hyperparameter space, reinforcing the importance of systematic tuning to achieve the best results.}

\

\newpage

\begin{table}[h]
\centering
\caption[]{Across cycle comparison of RMSE \cite{trained}  (Kelvin) of our probabilistic model with models based on Automatic Peak-detection (AP), Principal Component Regression (PCR), and 1D Convolutional Neural Networks (CNN)}

% \footnotemark (Kelvin) of our probabilistic model with models based on Automatic Peak-detection (AP), Principal Component Regression (PCR), and 1D Convolutional Neural Networks (CNN).}
\begin{tabular}{|c|c|c|c|c|c|c|}
\multicolumn{2}{c}{RMSE (Kelvin)} & \multicolumn{2}{c}{\uline{Expert Knowledge Driven}} & \multicolumn{2}{c}{\uline{Data Driven}} \\
\hline
Training & Testing &  Probabilistic Model & AP & PCR & CNN \\
\hline
Cycle2 & Cycle1 & 2.08 & 2.35 & {\bf 1.42} & 2.12 \\
\hline
Cycle1 & Cycle2 &  {\bf 1.47} & 3.14 & 15.3 & 18.75 \\
\hline
\end{tabular}
\label{tab: out sample}
\end{table}

\begin{table}[h]
\centering
\caption[Out-of-sample errors]{In cycle comparison of RMSE \cite{comparison} (Kelvin) of our probabilistic model with models based on Automatic Peak-detection (AP), Principal Component Regression (PCR), and 1D Convolutional Neural Networks (CNN).}
\begin{tabular}{|c|c|c|c|c|c|c|}
\multicolumn{2}{c}{RMSE (Kelvin)} & \multicolumn{2}{c}{\uline{Expert Knowledge Driven}} & \multicolumn{2}{c}{\uline{Data Driven}} \\
\hline
Training & Testing &  Probabilistic Model & AP & PCR & CNN \\
\hline
Cycle1 & Cycle1 & 1.9 & 2.14 & {\bf 1.23} & 2.16 \\
\hline
Cycle2 & Cycle2 & 1.7 & 2.84 & {\bf 1.59} & 1.75 \\
\hline
\end{tabular}
\label{tab:in sample}
\end{table}

\begin{table}[htbp]
\centering
\caption[]{Comparison of RMSE (Kelvin) of each model followed by a linear or quadratic residual model\cite{training}.}
\begin{tabular}{|c|c|c|c|c|c|}
\multicolumn{2}{c}{RMSE (Kelvin)} & \multicolumn{2}{c}{\uline{Expert Knowledge Driven}} & \multicolumn{2}{c}{\uline{Data Driven}} \\
\hline
Testing & Polynomial & Probabilistic Model & AP & PCR & CNN \\
\hline
Cycle1 & Linear & 1.11 & 2.3 & 1.16 & {\bf 1.03} \\
Cycle1 & Quadratic & 1.03 & 1.9 & 1.10 & {\bf 0.99} \\
\hline
Cycle2 & Linear & {\bf 1.62} & 2.83 & 5.65 & 4.24 \\
Cycle2 & Quadratic & {\bf 1.02} & 2.30 & 1.78 & 1.08 \\
\hline
\end{tabular}
\label{tab:residuals errors}
\end{table}

The trained model, after bias correction, outperforms the probabilistic model by 40 mK when tested within the training range (Table \ref{tab:residuals errors}). However, when required to extrapolate i.e., trained on Cycle 1 and tested on Cycle 2 the model performs nine times worse than the probabilistic model (see Tables \ref{tab: out sample} and \ref{tab:in sample}). This discrepancy arises because Cycle 2 spans a broader temperature range that the model has not encountered during training, unlike Cycle 1. As a result, the model struggles to generalize to these unseen temperatures, leading to significantly higher prediction errors.  \textcolor{black}{To mitigate this issue, one potential approach is to develop hybrid or ensemble models that combine data-driven techniques with probabilistic methods, leveraging the strengths of both frameworks. Another strategy involves utilizing Bayesian models, which provide uncertainty estimates alongside predictions, thereby enhancing the reliability of measurement outputs. Furthermore, Physics-Informed Neural Networks (PINNs) \cite{raissi2019physics} present a compelling alternative, as they can integrate probabilistic models to guide feature learning and improve generalization beyond the training range. Investigating these approaches offers a promising direction for future research aimed at enhancing the robustness and accuracy of data-driven sensor models.}

\section{Summary}
Successful adoption of emerging technologies in sensing, be they photonics or quantum-based, depends on a range of factors including cost, size, weight and power requirements, compatibility with existing infrastructure and workflow, and sensing performance metrics such as resolution and accuracy \cite{review2022}. In recent years researchers have demonstrated fiberization of photonic and quantum sensors can lead to fit-for-purpose devices that have similar dimensions as legacy sensors addressing an important concern of the user-community \cite{Janz:24,fiber_nv_mag}. Validating the performance metrics of packaged device including determining uncertainties is the focus of on-going work \cite{review2022, HARTINGS2019127076, Janz:24}. \\

\par In this study we have examined the impact the of inference models on uncertainties of a fiberized NV-diamond based temperature sensor by evaluating the model prediction uncertainties. We have compared the performance of a physics-informed probabilistic model with a conventional peak tracking methodology that draws on minimal domain knowledge, to data-driven models that are agnostic to the device physics. Our results demonstrate that data-driven models are easy to train, generalize well and within the training range, outperform the probabilistic model.   However, these models are data hungry and can be sensitive to random measurement noise particularly when the training datasets are small. Furthermore, whilst all models struggle with extrapolation, data-driven models performed worse, resulting in prediction uncertainties that were a factor of $\approx 10$ worse than the probabilistic model. \textcolor{black}{As shown in Tables \ref{tab: out sample} and \ref{tab:in sample}, the probabilistic model achieves an RMSE of 1.47 K when tested outside the training range, while PCR and CNN exhibit much higher uncertainties (15.3 K and 18.75 K, respectively). CNN performance improves post bias removal with additional modeling, suggesting that a more sophisticated architecture may help, though we avoided complexity to prevent overfitting. These results highlight the trade-offs between models: the probabilistic approach provides reliable uncertainty quantification and better extrapolation, while data-driven models leverage spectral features to enhance accuracy within the training range}

The unique strengths and weaknesses of probabilistic and data-driven models open new avenues for exploration. \textcolor{black}{The probabilistic model's adaptability allows it to be tailored for specific measurement scenarios, such as adjusting the Hamiltonian in the presence of a magnetic field or replacing the Gaussian noise model with a Poisson model for single NV measurements. Furthermore, as our understanding of the spectroscopic-temperature relationship improves, probabilistic models can be refined to enhance their accuracy.}

The transparency and uncertainty quantification of probabilistic models make them ideal for sensor fusion applications, such as Kalman filters. In contrast, data-driven models excel within the training range but face challenges when extrapolating beyond it. This suggests that a hybrid approach that combines the strengths of both models could provide the best performance. 

\textcolor{black}{Our work demonstrates that by leveraging the full spectral information, physics-informed probabilistic models can significantly outperform traditional parametric models, such as the auto peak finder algorithm. These models can streamline data processing, remove subjectivity, and deliver improved uncertainties. The observed model inference uncertainties (\(\pm 1\) K) indicate that cw-ODMR-based NV thermometry performs comparably to industrial thermometry technologies like thermocouples and fiber Bragg gratings\cite{review2022}. If further uncertainty reduction proves difficult, NV thermometry could be most valuable in niche applications, such as embedded sensors for quantum information systems, microfluidic systems, and biological imaging, where its small size, non-toxicity, and compatibility with existing protocols provide key advantages}.

\section{Acknowledgments:}
The authors would like to acknowledge AFMETCAL (R24-685-0005) for funding.  Shraddha Rajpal and Tyrus Berry would also like to acknowledge support from NSF grant DMS-2006808. 

\section{Data availability statement }
Data available upon reasonable request.

\bibliography{references}

\begin{thebibliography}{10}
\bibitem{range}
\enquote{Over wider temperature range e.g. 4 k to 600 k, the zero-field
  splitting, $d$, shows a complex non-linear dependence on temperature which
  can be modeled using a polynomial expansion,} .
\bibitem{wider}
\enquote{We note, that over wider temperature ranges it is known that that
  amplitude (or contrast) shows a modest inverse relationship whilst width
  shows a modest positive correlation. incorporating this relationship into the
  model is the subject of future work.} .
\bibitem{regularization}
\enquote{Imposing a l1 regularization on pca did not result in any significant
  improvement in performance (data not shown). examination of the top 3
  eigenmodes, shown in fig~\ref{fig:Modes} , reveals that the non-parametric
  model learns the temperature dependent changes in the spectral lineshape's
  moments including changes in peak center, width and amplitude. we note that
  the non-parametric approach is data hungry, as illustrated in
  fig~\ref{fig:Modes}, reduction in temperature range degrades the quality of
  the third mode, returning a mode dominated by random noise. as such the
  method reports factor of $\approx 10$ increase in prediction errors when
  tasked with extrapolation beyond the training range (table \ref{tab: out
  sample} .),} .
\bibitem{trained}
\enquote{Each method is trained/calibrated on one of two temperature cycles and
  tested on the other, the table reports the errors measured in the testing
  cycle, where cycle 2 is 243.14 k to 323.15 k and cycle 1 is 283.5 k to 323.15
  k. note that when trained on the wide temperature range (cycle 2) the data
  driven methods (pcr/cnn) have comparable performance to expert knowledge
  driven methods (top row). however, when trained on a narrow temperature range
  (cycle 1), the data driven methods struggle to generalize beyond the training
  range (bottom row).} .
\bibitem{comparison}
\enquote{Each method is trained/calibrated on one of two temperature cycles and
  then tested on out-of-sample data from the same cycle. when trained and
  tested on the same temperature range, the data-driven methods can even
  outperform the probabilistic model,} .
\bibitem{training}
\enquote{After each model is trained, the model is applied to the training data
  and then a linear or quadratic regression is fit between the model output and
  true temperatures. this same linear or quadratic function is then applied to
  the model output temperatures on the testing data. fitting these residuals
  reveals that a bias is introduced by the linear model, indicating that a
  non-linear or quadratic model may be necessary. the selection of the model is
  likely to benefit from a larger dataset. the rmses in this chart give the
  standard deviations between the data points and corresponding linear and
  quadratic best fit curves shown in figures \ref{fig:cycle2fit},
  \ref{fig:PCAtrain2test1}, \ref{fig:Peaktrain2test1}, and
  \ref{fig:NNtrain2test1}.} .
\end{thebibliography}

\end{document}